\documentclass[prd,aps,showpacs,preprintnumbers,amssymb]{revtex4}
\usepackage{graphicx}
\usepackage{epsf}
\usepackage{amsmath}
\usepackage{epstopdf}
\usepackage{multirow}
\usepackage{bm}
\def\e3p{$\eta \rightarrow 3 \pi$}

\begin{document}
\title{%
\hfill{\normalsize\vbox{%
\hbox{}
 }}\\
{Generalized linear sigma model with two glueballs}}

\author{Amir H. Fariborz
$^{\it \bf a}$~\footnote[1]{Email:
 fariboa@sunyit.edu}}

\author{Renata Jora
$^{\it \bf b}$~\footnote[2]{Email:
 rjora@theory.nipne.ro}}

\affiliation{$^{\bf \it a}$ Department of Matemathics/Physics, SUNY Polytechnic Institute, Utica, NY 13502, USA}
\affiliation{$^{\bf \it b}$ National Institute of Physics and Nuclear Engineering PO Box MG-6, Bucharest-Magurele, Romania}

\date{\today}

\begin{abstract}
	
We present a generalized linear sigma model that includes both scalar and pseudoscalar glueballs in addition to a quark-antiquark as well as a four-quark chiral nonet. Utilizing the axial and trace anomalies of QCD (at the effective mesonic level), we aim to develope the most general structure of the Lagrangian which can be used to study the interaction of quarkonia with glueballs.  We then study the effect of scalar glueball on the vacuum of the model by considering a decoupling limit in which the glueball fields are decoupled from quarkonia.  This determines the properties of the pure scalar glueball and builds a practical foundation for determining the model parameters when the interactions are turned on.

\end{abstract}
\maketitle

\section{Introduction}

Quantum Chromodynamics (QCD), the theory of  strong interactions \cite{PDG} displays in the low energy regime phenomena such as mass gap, confinement or approximate chiral symmetry.   Different attempts have been made to describe these properties (each with its own balance  of rigor versus practicality) including chiral perturbation theory \cite{ChPT1} and its extensions such as chiral unitary approach \cite{ChUA1}-\cite{ChUA8} and inverse amplitude method \cite{IAM1}-\cite{IAM3},   lattice QCD approaches \cite{LQCD1}-\cite{McNeile:2000xx}, QCD sum-rules \cite{SVZ}-\cite{Huang}, linear sigma models \cite{Schechter1}-\cite{Giacosa22}, as well as other explorations and non-perturbative methods \cite{Ioffe}-\cite{Mathieu}. At low energies the main degrees of freedom are mesons and baryons, bound states of two, three or more quarks or glueballs which are gauge invariant bound states of gluons with different possible quantum numbers \cite{Fritsch}-\cite{Mathieu}.

The possible presence of the glueballs in the QCD spectrum was discussed early on in \cite{Fritsch}. In the years that followed,  the glueball spectrum and properties  were analyzed in the context of QCD sum rules, quark constituent models or lattice QCD. For example in quenched lattice approximation the mass of the lowest scalar glueball was calculated to be  $M_{0+}=1.550$ GeV \cite{Bali1}, $M_{0+}=1.730$ GeV \cite{Morningstar} or $M_{0+}=1.709$ GeV in \cite{Chen} (see \cite{Ochs} for a thorough review). In the QCD sum rules approach,  the lowest scalar glueballs are predicted to be \cite{Narison22}-\cite{Narison3} $M_{0^+}=0.9-1.1$ GeV and $M'_{0^+}=1.5-1.6$ GeV with the possibility of   a broad lower state $M_{0+}=0.7$ GeV (\cite{Narison22,Narison3}).    However,  experiment has not yet detected ``pure'' scalar (or pseudoscalar) glueballs, because, expectedly,
these composites mix with the appropriate quark building blocks of mesons and thereby ``hide'' inside some of the known mesonic states such as several of the isosinglet scalar states in the 1-2 GeV range [$f_0(1370)$, $f_0(1500)$ and $f_0(1710)$] which are considered to have noticeable glue admixtures.

For the lowest pseudoscalar glueball states there are many candidates  proposed in different frameworks. One of them is 
$\eta(1405)$ \cite{Scharre} while other candidates below $2$ GeV proposed in  \cite{Mathieu}.   
The observation that eta states around 1.5 GeV are more complex states than a simple quark-antiquark  is shared by other approaches such as the work of  \cite{ChUA8} which has studied the possibility of  these states being dynamically generated in $\eta f_0(980)$ and $\pi a_0(980)$ interactions.  In our investigation too \cite{Jora5},  the eta states around 1.5 GeV are not pure (or dominantly) quark-antiquark states and their compositions seem to contain a large four-quark and glue admixtures.
In lattice QCD the mass of the lowest pseudoscalar glueball is estimated at $M_{0^-}=2.330$ GeV in \cite{Bali1}, $M_{0^-}=2.590$ GeV in \cite{Morningstar} and $M_{0^-}=2.557$ GeV in \cite{Chen}.  Similar to the scalar glueball mixing with scalar isosinglet quark states, the quark bound states with the same quantum numbers mix with the pseudoscalar glueball states leading to a complex spectroscopy worthy of investigation. For example the interference between the glueball states and the quark-antiquark ones in the lattice method was studied in \cite{Hart1}, \cite{Hart2} with the main conclusion that there is a maximal mixing between the glueball states and two quark meson singlet ones.

In the absence of a fully understood theoretical framework for low-energy QCD, determination
of various quark and glue components of the physical states  is known
to be notoriously difficult and any attempt on disentanglement of such components within a
given framework inevitably suffers from model dependencies.     Such model dependencies can be minimized if the framework is tested against various low-energy processes.
In \cite{Jora1}-\cite{Jora7} we proposed and developed a generalized linear sigma model with two chiral nonets  that satisfy approximately  the low-energy chiral symmetry and the quantum anomalies of QCD.  The model was able to describe the mass spectrum and some of the properties of $36$ low lying scalar and pseudoscalar states with a good agreement with the experimental data.  The axial and trace anomalies in a linear sigma model refer both to the electromagnetic and the gluon fields. In \cite{Jora6,Jora7} we studied the effect of the electromagnetic axial and trace anomalies on the decays to two photons of some of the scalar and pseudoscalar states. However the treatment of the gluon anomalies is somewhat different because the gauge invariant glueball states may behave as individual degrees of freedom which may be either integrated out from the Lagrangian or maintained as physical states. The purpose of our work will thus be to investigate how the low lying physical scalar and pseudoscalar glueballs may fit in a generalized linear sigma model with two chiral nonets, one with a quark-antiquark structure, the other one with a four quark composition.

This work is organized as follows:   In Sec. II, we give the main templates for modeling the quantum anomalies of QCD at the mesonic level and establish connections with the underlying fundamental properties.    For the convenience of the readers, in Sec. III, we give a brief review of our generalized linear sigma model (in the absence of glueballs),  and then show how glueballs can be added to this framework in Sec. IV.    This leads to an extended version of the  generalized linear sigma model, which naturally comes with the price of additional complexities due to the proliferation of  new parameters.    Tackling this Lagrangian requires a careful and  ground-up approach in which addition of glueballs and their interactions with quarkonia components are tractable.   This brings us to Sec. V where we consider a decoupling limit in which  glueballs, while present in the model, do not interact with quark composite operators,  and thus allowing a probe of their role in stabilizing the QCD vacuum and measuring the direct effect of glueball condensate on the model parameters.  This section serves as a foundation for further studies when the interactions are turned on.    Additional relationships and bulkier formulas are collected in two appendices.

\section{ A two glueball Lagrangian}

In \cite{Schechter} Schechter proposed an effective Lagrangian that contains two glueball states; a pseudoscalar glueball that satisfies the $U(1)_A$ anomaly and a scalar glueball that satisfies the trace anomaly according to:
\begin{eqnarray}
&&\partial^{\mu}J^5_{\mu}=\frac{g^2}{16\pi^2}N_F\tilde{F}F=G
\nonumber\\
&&\theta^{\mu}_{\mu}=\partial^{\mu}D_{\mu} = -\frac{\beta(g^2)}{2g}FF=H.
\label{intr64553}
\end{eqnarray}
Here $F$ is the $SU(3)_C$ field tensor, $\tilde{F}$ is its dual, $N_F$ is the number of flavors, $\beta(g^2)$ is the beta function for the coupling constant, $J^5_{\mu}$ is the axial current and $D_{\mu}$ is the dilatation current.

Assuming that the two glueballs are not physical states but are integrated out by using the equation of motion we can derive the following Lagrangian that satisfies the axial and trace anomalies:

\begin{eqnarray}
{\cal L}=-\frac{1}{2}{\rm Tr}(\partial^{\mu}M\partial_{\mu}M^{\dagger})+f(I_n,G,H)
+\frac{i}{4N_f}G\,
\ln\left(
{{ \det M} \over{\det M^{\dagger}}}
\right)
-H\,\sum_m\frac{c_m}{m}\ln\left(\frac{R_m}{\Lambda^m}\right).
\label{firstlagr665}
\end{eqnarray}

Here, $\sum_m c_m = 1$,  $M$ is the two quark chiral nonet field of pseudoscalar and scalar states,  $I_n={\rm Tr}(MM^{\dagger})^n$, and  $f(I_n,G,H)=f(I_n,-G,H)$ is in general a chiral and U(1)$_{\rm A}$ invariant function that must satisfy the scale invariance condition:
\begin{eqnarray}
{\rm Tr}\left[M\frac{\partial f}{\partial M} +  M^{\dagger}   \frac{\partial f}{\partial M^{\dagger}}\right]+4G\frac{\partial f}{\partial G}+4H\frac{\partial f}{\partial H}=4f.
\label{cnstr664554}
\end{eqnarray}
Equivalently, since $f$ is chiral-invariant and hence a function of $I_n$, its scale invariance also implies:
\begin{eqnarray}
\sum_n	2 n I_n\frac{\partial f}{\partial I_n}+4G\frac{\partial f}{\partial G}+4H\frac{\partial f}{\partial H}=4f.
	\label{finv_In}
\end{eqnarray}
This partial differential equation can be solved to give the general form of function $f$:
\begin{equation}
f(I_n,G,H) = \sum_{i} k_i
\left(
\prod_n I_n^{(l_n)_i}
\right)
 G^{2p_i}  H^{q_i}
\end{equation}
where $k_i$ are unknown constants and $\sum_n 2n \left(l_n\right)_i + 8p_i + 4q_i = 4$.    The leading terms in $f$ up to  quadratic power of the fields are:
\begin{equation}
  {\sqrt H} I_1,  I_2, H, {G^2\over I_2},  {G^2\over H}, {H^2\over I_2},  \cdots
\label{f_n_expan}
\end{equation}
Note that terms such as $I_1^2$,  $I_2^2/H$, $G^2 H/I_2^2$ involve two separate flavor traces and are not favored by OZI rule and therefore are not as important as the above terms.   In addition,  terms that include $I_2^2$ have higher number of quark and antiquark lines and according to the approximation scheme developed in \cite{Jora5} (which is one of the guiding criteria of the present work) are considered less important compared to terms with linear power of $I_2$.

Similarly,  $R_m$ is an arbitrary function of $I_n$, $G$ and $H$ that must satisfy the requirements:
\begin{eqnarray}
&&{\rm Tr}\left[M\frac{\partial R_m}{\partial M}+\frac{\partial R_m}{\partial M^{\dagger}}M^{\dagger}\right]+4G\frac{\partial R_m}{\partial G}+4H\frac{\partial R_m}{\partial H}=mR_m
\nonumber\\
&&R_m(I_n,G,H)=R_m(I_n,-G,H).
\label{secset45364554}
\end{eqnarray}
Again, the partial differential equation for $R_m$ can be rewritten in terms of $I_n$,
\begin{equation}
\sum_n 2 n I_n\frac{\partial R_m}{\partial I_n}+4G\frac{\partial R_m}{\partial G}+4H\frac{\partial R_m}{\partial H}=mR_m
\label{Inv_Rm_In}
\end{equation}
with the general solution
\begin{equation}
R_m(I_n,G,H) = \sum_{i} r_i^{(m)}
\left(
\prod_n I_n^{(l_n)_i}
\right)
G^{2p_i}  H^{q_i}
\end{equation}
where $r_i^{(m)}$ are unknown constants and $\sum_n 2n (l_n)_i + 8 p_i + 4 q_i = m$.

Terms contributing to  $m=2$ include,
\begin{equation}
I_1,  {I_2\over I_1},   {H\over I_1},  \cdots
\label{R2_expan}
\end{equation}
Terms with $m=3$ include,
\begin{equation}
\det M,  \det M^\dagger,    \cdots
\label{R3_expan}
\end{equation}
Terms contributing to $R_4$ are the same as (\ref{f_n_expan}), etc.

Our objective in this work is to use the same methodology as presented in this section to extend the generalized linear sigma model that contains also a four-quark chiral nonet $M'$ in addition to the quark-antiquark chiral nonet $M$ present in the above formulation.

\section{Brief review of the generalized linear sigma model}


The model is constructed in terms of 3$\times$3 matrix
chiral nonet fields:
\begin{equation}
M = S +i\phi, \hskip 2cm
M^\prime = S^\prime +i\phi^\prime,
\label{sandphi1}
\end{equation}
which are in turn defined in terms of ``bare'' scalar meson nonets $S$ (a quark-antiquark scalar nonet) and $S'$ (a four-quark scalar nonet), as well as ``bare'' pseudoscalar meson nonets $\phi$ (a quark-antiquark pseudoscalar nonet) and $\phi'$ (a four-quark pseudoscalar nonet).
Chiral fields $M$ and $M'$ transform in the same way under
chiral SU(3) transformations
\begin{eqnarray}
M &\rightarrow& U_L\, M \, U_R^\dagger,\nonumber\\
M' &\rightarrow& U_L\, M' \, U_R^\dagger,
\end{eqnarray}
but transform differently under U(1)$_A$
transformation properties
\begin{eqnarray}
M &\rightarrow& e^{2i\nu}\, M,  \nonumber\\
M' &\rightarrow& e^{-4i\nu}\, M'.
\label{U1A}
\end{eqnarray}
There are several possible four-quark substructures for $M'$ (such as diquark-antidiquark types  or molecular type), however, the model does not distinguish these different types of four-quark substructures and can only probe the percentages of quark-antiquark and four-quark components (but not different types of four-quark components).    The model distinguishes  $M$  from $M'$ through the U(1)$_A$ transformation according to (\ref{U1A}).

The Lagrangian density has the general structure
\begin{equation}
{\cal L} = - \frac{1}{2} {\rm Tr}
\left( \partial_\mu M \partial_\mu M^\dagger
\right) - \frac{1}{2} {\rm Tr}
\left( \partial_\mu M^\prime \partial_\mu M^{\prime \dagger} \right)
- V_0 \left( M, M^\prime \right) - V_{SB},
\label{mixingLsMLag}
\end{equation}
where $V_0(M,M^\prime)$ stands for a function made
from SU(3)$_{\rm L} \times$ SU(3)$_{\rm R}$
(but not necessarily U(1)$_{\rm A}$) invariants
formed out of $M$ and $M^\prime$.    In addition to scalar and pseudoscalar mesons included in this Lagrangian density,  the vector and axial vector mesons can be introduced by gauging the linear sigma model \cite{GaugedLSM}.  However, for investigation of the scalar and pseudoscalar mass spectrum (which is the main objective of the present work) inclusion of vectors and axia vectors are of qualitatively limited relevance. 
In principle,  there are infinite number of invariant  terms in the potential.   To keep the calculations in this model tractable, it is practical to define an approximation scheme
that allows limiting the number of terms at each level of calculation, and systematically improving the results thereafter.
Such a scheme was defined in \cite{Jora41}, in terms of the number of underlying quark and antiquark fields in each term.
The  leading choice of terms
corresponding
to eight or fewer underlying quark plus antiquark lines
at each effective vertex
reads:
\begin{eqnarray}
V_0 =&-&c_2 \, {\rm Tr} (MM^{\dagger}) +
c_4^a \, {\rm Tr} (MM^{\dagger}MM^{\dagger})
\nonumber \\
&+& d_2 \,
{\rm Tr} (M^{\prime}M^{\prime\dagger})
+ e_3^a(\epsilon_{abc}\epsilon^{def}M^a_dM^b_eM'^c_f + {\rm H. c.})
\nonumber \\
&+&  c_3\left[ \gamma_1 {\rm ln} (\frac{{\rm det} M}{{\rm det}
	M^{\dagger}})
+(1-\gamma_1){\rm ln}\frac{{\rm Tr}(MM'^\dagger)}{{\rm
		Tr}(M'M^\dagger)}\right]^2.
\label{SpecLag}
\end{eqnarray}
All the terms except the last two (which mock up the axial anomaly)
have been chosen to also
possess the  U(1)$_{\rm A}$
invariance.   The symmetry breaking term which models the QCD mass term
takes the form:
\begin{equation}
V_{SB} = - 2\, {\rm Tr} (A\, S),
\label{vsb}
\end{equation}
where $A={\rm diag} (A_1,A_2,A_3)$ are proportional to
the three light quark
current masses (i.e. in the isospin invariant limit $A_1=A_2 \propto m_u=m_d$ and $A_3\propto m_s$.)
The model allows for two- and four-quark condensates,
\begin{eqnarray}
\alpha_a &=& \langle S_a^a \rangle, \nonumber\\
\beta_a  &=& \langle {S'}_a^a \rangle.
\label{condensates}
\end{eqnarray}
Here we assume isotopic spin
symmetry so A$_1$ =A$_2$ and:
\begin{equation}
\alpha_1 = \alpha_2  \ne \alpha_3, \hskip 2cm
\beta_1 = \beta_2  \ne \beta_3.
\label{ispinvac}
\end{equation}
We also need the ``minimum" conditions,
\begin{equation}
\left< \frac{\partial V_0}{\partial S}\right>_0 + \left< \frac{\partial
	V_{SB}}{\partial
	S}\right>_0=0,
\quad \quad \left< \frac{\partial V_0}{\partial S'}\right>_0
=0,
\label{mincond}
\end{equation}
where the brackets with subscript zero represent evaluation of the derivatives at the vacuum expectation values (\ref{condensates}).
At the leading order of the model (containing  terms with eight or fewer quark and antiquark lines) there are twelve parameters describing the Lagrangian and the vacuum. These include the six coupling constants
given in Eq. (\ref{SpecLag}), the two quark mass parameters,
($A_1=A_2,A_3$) and the four vacuum parameters ($\alpha_1
=\alpha_2,\alpha_3,\beta_1=\beta_2,\beta_3$). The four minimum
equations reduce the number of needed input parameters to
eight.   In the work of \cite{Jora5},  these eight experimental inputs were selected from several masses of relevant states together with pion decay constant and  the light
``quark mass ratio" $A_3/A_1$, allowing a complete determination of the Lagrangian parameters and making  predictions for some of the unknown masses and two and four-quark percentages.    It was found that there is a significant underlying mixings among the two- and   four-quark components of scalars below and above 1 GeV with the four-quark components of  those below 1 GeV having an edge over their quark-antiquark components.   This is in contrast to the physical light pseudoscalar meson nonet below 1 GeV for which this picture is reversed.     Inclusion of both scalar and pseudoscalar glueballs is expected to improve this analysis.   This directly affects the properties of isosinglet states, which in addition to two- and four-quark components, can contain a glue content.   While the case of isodoublets and isotriplets are not directly affected by the inclusion of glueballs, however, when glueballs are included in this model, they can generally modify the model parameters and thereby can indirectly affect the properties of these states as well.


\section{Inclusion of scalar and pseudoscalar glueballs in the generalized linear sigma model}

The formal extension to two chiral nonets $M$ and $M'$ (which is the main focus of this work) is straightforward:

\begin{equation}
{\cal L} = -\frac{1}{2} {\rm Tr}(\partial^{\mu}M\partial_{\mu}M^{\dagger})-\frac{1}{2} {\rm Tr}(\partial^{\mu}M'\partial_{\mu}{M'}^{\dagger}) - \frac{1}{32} H^{-3/2}\partial^{\mu}H\partial_{\mu}H  - \frac{1}{2}H^{-3/2}\partial^{\mu}G\partial_{\mu}G  + f + f_{\rm A} + f_{\rm S} + f_{SB}.
\label{inlgr56}
\end{equation}
Here $f$ is invariant under the chiral symmetry and $U(1)_A$, $f_{\rm A}$ is a term that mocks up the axial anomaly, $f_{\rm S}$ is a term that leads to the correct scale anomaly and $f_{SB}$ introduces explict breaking of the chiral symmetry in the Lagrangian. In what follows we will discuss in detail the properties and  expressions for each of these terms.

The first term after kinetic terms is $f$ with the general form  $f(I_n, I'_m, I''_{st}, G,H) = f(I_n, I'_m, I''_{st},-G,H)$ where $I_n={\rm Tr}\left[ \left(M M^{\dagger}\right)^n\right]$,
$I'_m={\rm Tr}\left[ \left(M' {M'}^{\dagger}\right)^m\right]$,
$I''_{st}$ is a hermitian and chiral invariant combination that contains $s$ number of fields $M$ and $t$ number of  fields $M'$ and thus has the mass dimension $s+t$ (note that for each pair $s$ and $t$ there are multiple possibilities for $I_{st}''$ which are however encapsulated in the same abstract  notation for simplicity). We require the invariance of $f$ under the scale transformation according to:  
\begin{eqnarray}
{\rm Tr}\left[M\frac{\partial f}{\partial M}+ M^{\dagger} \frac{\partial f}{\partial M^{\dagger}}\right] +
{\rm Tr}\left[M'\frac{\partial f}{\partial M'}+ {M'}^{\dagger}\frac{\partial f}{\partial {M'}^{\dagger}}\right]
+4G\frac{\partial f}{\partial G}+4H\frac{\partial f}{\partial H}=4f.
\label{cnstr_inv}
\end{eqnarray}
Equivalently, since $f$ is chiral-invariant and hence a function of $I_n$, $I'_n$ and $I''_{nm}$ its scale invariance also implies:
\begin{eqnarray}
\sum_n	2 n I_n\frac{\partial f}{\partial I_n} +
\sum_m	2 m I'_m\frac{\partial f}{\partial I'_m} +
\sum_{st}	(s+t) I''_{st}\frac{\partial f}{\partial I''_{st}}
+ 4G\frac{\partial f}{\partial G}+4H\frac{\partial f}{\partial H}=4f.
\label{finv_In2}
\end{eqnarray}
This partial differential equation can be solved to give the general form of function $f$:

\begin{equation}
f (I_n, I'_n, I''_{nm}, G,H) = \sum_i u_i
\left(
\prod_n I_n^{(l_n)_i}
\right)
\left(
\prod_m {I'}_m^{(l_m)_i}
\right)
\left(
\prod_{st} {I''}_{st}^{(l_{st})_i}
\right)
 G^{2p_i}  H^{q_i}
\label{E_f_general}
\end{equation}
where $u_i$ are unknown constants and $\sum_n 2n \left(l_n\right)_i +  \sum_m 2m \left(l_m\right)_i + \sum_{st} (s + t) \left({l}_{st}\right)_i + 8p_i + 4q_i = 4$.    The leading terms in $f$ up to  quadratic power of the fields are:
\begin{equation}
{\sqrt H} I_1,  I_2,  {\sqrt H} I'_1,     H, {G^2\over I_2},  {G^2\over I'_2}, {G^2\over H}, {H^2\over I_2},  {H^2\over I'_2} \cdots
\label{f_n_expan}
\end{equation}
As stated in Sec. II, again note that terms such as $I_1^2$,  $I_2^2/H$, $G^2 H/I_2^2$ include separate flavor traces and are not favored by the OZI rule. Also terms such as $I_2^2$ contain  higher number of quark and antiquark lines and thus less important compared to terms like $I_2$ with fewer lines.

The term corresponding to the axial anomaly must have the general expression:
\begin{equation}
f_{\rm A} =
i{G\over 12} \sum_i \gamma_i \ln \left( Q_i\over Q_i^\dagger\right)
\label{potential73568}
\end{equation}
where $\sum_i \gamma_i = 1$ and composite operators $Q_i$ can be selected from the set
\begin{equation}
Q_i \in \left\{ \det M, \det M', {\rm Tr} \left(M {M'}^\dagger \right), \epsilon_{abc}\epsilon^{def}M_d^a {M'}_e^b{M'}_f^c,
 {\rm Tr} \left(M {M'}^\dagger M {M'}^\dagger \right), \cdots     \right\}
\label{E_Qi_set}
\end{equation}

The term leading to the correct scale anomaly is given by:
\begin{equation}
f_{\rm S} =
- H \sum_m \tau_m \ln \left( R_m \over \Lambda^m\right),
\label{potential7356}
\end{equation}
where the arbitrary parameters $\tau_m$ must satisfy the constraint: $\sum_mm\tau_m=1$.

Similarly,  $R_m$ is an arbitrary function of $I_n$, $I'_m$,$I''_{st}$, $G$ and $H$ that must satisfy the requirements:
\begin{eqnarray}
&&{\rm Tr}\left[M\frac{\partial R_m}{\partial M}+   M^{\dagger}   \frac{\partial R_m}{\partial M^{\dagger}} \right] +
{\rm Tr}\left[M'\frac{\partial R_m}{\partial M'}+  {M'}^{\dagger}\frac{\partial R_m}{\partial {M'}^{\dagger}}\right]
+4G\frac{\partial R_m}{\partial G}+4H\frac{\partial R_m}{\partial H}=mR_m
\nonumber\\
&&
R_m(I_n, I'_m, I''_{st}, G,H) = R_m(I_n, I'_m, I''_{st},-G,H)
\label{secset45364554}
\end{eqnarray}
Again, the partial differential equation for $R_m$ can be rewritten in terms of $I_n$, $I'_m$ and $I''_{st}$
\begin{equation}
\sum_n 2 n I_n\frac{\partial R_m}{\partial I_n}+ \sum_m 2 m I'_m\frac{\partial R_m}{\partial I'_m}+
\sum_{st} (s + t) I''_{st}\frac{\partial R_m}{\partial I''_{st}}+
4G\frac{\partial R_m}{\partial G}+4H\frac{\partial R_m}{\partial H}=mR_m
\label{Inv_Rm_In}
\end{equation}
with the general solution
\begin{equation}
R_m (I_n, I'_n, I''_{nm}, G,H) = \sum_i v_i^{(m)}
\left(
\prod_n I_n^{(l_n)_i}
\right)
\left(
\prod_m {I'}_m^{(l_m)_i}
\right)
\left(
\prod_{st} {I''}_{st}^{(l_{st})_i}
\right)
G^{2p_i}  H^{q_i}
\end{equation}
where $v_i^{(m)}$ are unknown constants and $\sum_n 2n \left(l_n\right)_i +  \sum_m 2m \left(l_m\right)_i + \sum_{st} (s + t) \left({l}_{st}\right)_i + 8p_i + 4q_i = m$.

   Terms contributing to  $m=2$ include,
\begin{equation}
I_1,  {I_2\over I_1},   {H\over I_1},  \cdots
\label{R2_expan}
\end{equation}
Terms with $m=3$ include,
\begin{equation}
\det M,  \det M^\dagger,    \cdots
\label{R3_expan}
\end{equation}
Terms contributing to $R_4$ are the same as (\ref{f_n_expan}), etc.

Finally there are multiple possibilities for the symmetry breaking term $f_{SB}$ \cite{Jora1}. An explicit example will be given at the end of this section.

The scalar and pseudoscalar fields $H$ and $G$ (that have mass dimension 4) are related to  scalar and pseudoscalar fields $h$ and $g$ with mass dimension 1. We make the substitution $H=h^4$ and $G=h^3g$ in the Lagrangian (\ref{inlgr56}) to obtain:

\begin{eqnarray}
{\cal L}&=&-\frac{1}{2}{\rm Tr}(\partial^{\mu}M\partial_{\mu}M^{\dagger})-\frac{1}{2}{\rm Tr}(\partial^{\mu}M'\partial_{\mu}{M'}^{\dagger})
-{1\over 2} (\partial_{\mu} h)(\partial_{\mu} h) - {1\over 2} (\partial_{\mu} g) (\partial_{\mu} g)
-V, \nonumber \\
- V&=& f + f_{\rm A} + f_{\rm S} + f_{\rm SB}.
\label{inlgr567}
\end{eqnarray}
where, as discussed previously,  $f(M, M', g, h)$ is invariant under chiral, axial and scale transformations, whereas $f_{\rm A}$ and $f_{\rm S}$ respectively break axial and scale symmetries according to (\ref{intr64553}) and $f_{\rm SB}$ is explicit symmetry breaker due to quark masses.

Below, we give the particular expressions for $f$, $f_{\rm A}$ and $f_{\rm S}$ such that the inclusion of glueballs is achieved as the minimal extension of the leading order of generalized linear sigma model discussed in Sec. III. Since the resulting mass spectra and probe of the substructures are extracted from a set of highly nonlinear and coupled system of equations,   it is important to first carefully study this minimal extension in order to keep the calculations manageable and be able to provide a meaningful interpretation of the results in comparison with those found previously in \cite{Jora5} (and references therein).

In the minimal extension,  the part of function $f$ that contains quark-antiquark and four-quark chiral nonets $M$ and $M'$ corresponds to terms in the general expression for $f$ [given in (\ref{E_f_general})] that recover the first four terms of potential (\ref{SpecLag}).  Since function $f$ is also scale invariant, effectively the desired extension is obtained  by replacing the dimensionful couplings in the first four terms of (\ref{SpecLag}) with dimensionless couplings and appropriate powers of scalar glueball field $h$, i.e. by making the following substituation in (\ref{SpecLag})
\begin{eqnarray}
c_2 &\rightarrow& - u_1 h^2, \nonumber \\
c_4^a &\rightarrow&  u_4,  \nonumber \\
d_2 &\rightarrow&  u_3 h^2, \nonumber \\
e_3^a &\rightarrow&  u_4 h,  
\label{E_f_subs}
\end{eqnarray}
which, when combined with the mass terms for scalar and pseudoscalar glueballs, result in the following chiral, U(1)$_{\rm A}$, and scale invariant terms in $f$ in this minimal extension        
\begin{eqnarray}
f(M, M', g, h) &=&
- \left(
u_1 h^2 {\rm Tr}[MM^{\dagger}]
+ u_2{\rm Tr}[MM^{\dagger}MM^{\dagger}]+
u_3 h^2 {\rm Tr}[M^{\prime}M^{\prime \dagger}]+ u_4 h (\epsilon_{abc}\epsilon^{def}M^a_dM^b_eM^{\prime c}_f+h.c.)+
\right.
\nonumber \\
&&
\left.\hskip .5cm  u_5 h^4 + u_6  h^2 g^2  + \cdots\right).
\label{pot201867}
\end{eqnarray}
Similarly,  we can write down $f_{\rm A}$ in the minimal extension.   As mentioned in Sec. III,  in the leading order of generalized linear sigma model in which only effective terms with eight (or fewer) quark or antiquark lines are retained, the effective term for axial anomaly is the last term given in Eq. (\ref{SpecLag}) which is obtained from integrating the pseudoscalar glueball out.  When the pseudoscalar glueball is present in the Lagrangian,  the general form that mocks up the exact U(1)$_{\rm A}$ anomaly is given in (\ref{potential73568}) with operators $Q_i$ taken from set (\ref{E_Qi_set}).  In the minimal extension, the operators  that (after integrating out the pseudoscalar glueball field)  result in the last term of (\ref{SpecLag}) are $\det M$ and ${\rm Tr}\left( MM'^\dagger\right)$ (note that $\det M'$ contains 12 quark and aniquark lines and does not contribute to the leading order of generalized linear sigma model).    Therefore, in the minimal extension
\begin{eqnarray}
f_{\rm A} &=&
i{G\over 12} \left[ \gamma_1\ln \left(\frac{\det M}{\det M^{\dagger}}\right)+\gamma_2\ln\left(\frac{{\rm Tr}(MM^{\prime\dagger})}{{\rm Tr}(M^{\prime}M^{\dagger})}\right)\right],
\label{axial2018967}
\end{eqnarray}
where  $\gamma_1$ and $\gamma_2$ in Eq. (\ref{axial2018967}) are arbitrary parameters that must satisfy the constraint: $\gamma_1+\gamma_2=1$ \cite{Jora1}-\cite{Jora6}.    

In a similar fashion, we can work out $f_{\rm S}$ in this minimal extension using the general template (\ref{potential7356}). Incorporating the same operators $\det M$ and ${\rm Tr}\left( MM'^\dagger\right)$ that we just discussed above for the axial anomaly, as well as a term which is the fourth power of the glueball field $h$,  Eq. (\ref{potential7356}) results in:  
\begin{eqnarray}
f_{\rm S} &=&
-H \left\{
\lambda_1 \ln\left(\frac{H}{\Lambda^4}\right)
+\lambda_2 \left[\ln\left(\frac{\det M}{\Lambda^3}\right)+\ln\left(\frac{\det M^{\dagger}}{\Lambda^3}\right)\right]
\right.
\nonumber \\
&&
\left.
\hskip .8cm
+ \lambda_3\left[ \ln\left(\frac{{\rm Tr} MM^{\prime\dagger}}{\Lambda^2}\right)+\ln\left(\frac{{\rm Tr}M' M^{\dagger}}{\Lambda^2}\right)
\right]\right\},
\label{scale7756}
\end{eqnarray}
where  $\Lambda$ with mass dimension one is the characteristic scale of QCD and  $\lambda_1$, $\lambda_2$ and $\lambda_3$ are  arbitrary parameters that must fulfill the condition: $4\lambda_1+6\lambda_2+4\lambda_3=1$ \cite{Jora7}.
As such, the terms $f$  and $f_{\rm S}$ in the potential are invariant under U(3)$_{\rm L} \times$ U(3)$_{\rm R}$ and  $f_{\rm A}$  breaks U(1)$_{\rm A}$.

In the presence of the quark masses,
\begin{eqnarray}
\theta^{\mu}_{\mu}=H-\left(1+\gamma_m \right) V_{SB},
\label{gamma7775}
\end{eqnarray}
where $\gamma_m$ is the anomalous dimension of the fermion mass operator. 
Note that a simple symmetry breaking term such as (\ref{vsb}) does not fully result in Eq. (\ref{gamma7775}), therefore, it should be extended.  The complete symmetry breaking term is:
\begin{eqnarray}
f_{\rm SB}={\rm Tr} [MM^{\dagger}]^{1-\frac{\gamma_m}{2}}{\rm Tr}[A(M+M^{\dagger}].
\label{symbre77555}
\end{eqnarray}
which, under the scale transformation, leads to exactly  the second term on the right hand side of Eq. (\ref{gamma7775}), 
where, similar to (\ref{vsb}),  $A={\rm diag}(A_1,A_2,A_3)$ is proportional to the three light quark masses.

The minimum equations describing the stability of vacuum are
\begin{eqnarray}
\left\langle\frac{\partial V}{\partial S^1_1}\right\rangle_0&=&
4\,{\it u_4}\,{\it h_0}\, \left( \alpha_1\,\beta_3+\alpha_3\,\beta_1
\right) +2\,{\it u_1}\,{{\it h_0}}^{2}\alpha_1+4\,{\it u_2}\,{\alpha_1}^{3
}+4\,{\frac { \left( \beta_1\, \left( \lambda_2+\lambda_3/2 \right)
	\alpha_1+1/2\,\alpha_3\,\beta_3\,\lambda_2 \right) {{\it h_0}}^{4}}{2\,{
		\alpha_1}^{2}\beta_1+\alpha_1\,\alpha_3\,\beta_3}}
\nonumber\\
&&
+4\, \left( 2\,{\alpha_1}^
{2}+{\alpha_3}^{2} \right) ^{-{\it \gamma_m}/2} \left( -2+{\it
	\gamma_m} \right)  \left( {\it A_1}\,\alpha_1+1/2\,{\it A_3}\,\alpha_3
\right) \alpha_1-2\, \left( 2\,{\alpha_1}^{2}+{\alpha_3}^{2} \right) ^{1
	-{\it \gamma_m}/2}{\it A_1}
\nonumber\\
\left\langle\frac{\partial V}{\partial S^3_3}\right\rangle_0&=&
8\,{\it u_4}\,{\it h_0}\,\beta_1\,\alpha_1+2\,{\it u_1}\,{{\it h_0}}^{2}
\alpha_3+4\,{\it u_2}\,{\alpha_3}^{3}+4\,{\frac { \left( 1/2\,\beta_3\,
		\left( \lambda_2+\lambda_3 \right) \alpha_3+\alpha_1\,\beta_1\,\lambda_2
		\right) {{\it h_0}}^{4}}{2\,\alpha_1\,\alpha_3\,\beta_1+\beta_3\,{\alpha_3}
		^{2}}}
\nonumber \\
&&
+4\, \left( 2\,{\alpha_1}^{2}+{\alpha_3}^{2} \right) ^{-{\it
		\gamma_m}/2} \left( -2+{\it \gamma_m} \right) \alpha_3\, \left( {
	\it A_1}\,\alpha_1+1/2\,{\it A_3}\,\alpha_3 \right) -2\, \left( 2\,{\alpha_1}^{2}+{\alpha_3}^{2} \right) ^{1-{\it \gamma_m}/2}{\it A_3}
\nonumber\\
\left\langle\frac{\partial V}{\partial {S'}^1_1}\right\rangle_0&=&
2\,{\frac { \left( 4\,{\alpha_1}^{2}\alpha_3\,\beta_1\,{\it u_4}+ \left( 2
		\,{\alpha_3}^{2}\beta_3\,{\it u_4}+2\,{\beta_1}^{2}{\it h_0}\,{\it u_3}+{{
				\it h_0}}^{3}\lambda_3 \right) \alpha_1+\alpha_3\,\beta_1\,\beta_3\,{\it h_0}
		\,{\it u_3} \right) {\it h_0}}{2\,\beta_1\,\alpha_1+\beta_3\,\alpha_3}}
\nonumber \\
\left\langle\frac{\partial V}{\partial {S'}^3_3}\right\rangle_0&=&
2\,{\frac {{\it h_0}\, \left( 4\,{\alpha_1}^{3}\beta_1\,{\it u_4}+2\,{
			\alpha_1}^{2}\alpha_3\,\beta_3\,{\it u_4}+2\,\alpha_1\,\beta_1\,\beta_3\,{
			\it h_0}\,{\it u_3}+\alpha_3\,{\beta_3}^{2}{\it h_0}\,{\it u_3}+\alpha_3\,{{
				\it h_0}}^{3}\lambda_3 \right) }{2\,\beta_1\,\alpha_1+\beta_3\,\alpha_3}}
\nonumber \\
\left\langle\frac{\partial V}{\partial h}\right\rangle_0&=&
8\,\ln  \left( {\frac {2\,\beta_1\,\alpha_1+\beta_3\,\alpha_3}{{\Lambda}^{
			2}}} \right) {{\it h_0}}^{3}\lambda_3+8\,\ln  \left( {\frac {{\alpha_1}^{
			2}\alpha_3}{{\Lambda}^{3}}} \right) {{\it h_0}}^{3}\lambda_2+4\,\ln
\left( {\frac {{{\it h_0}}^{4}}{{\Lambda}^{4}}} \right) {{\it h_0}}^{3}
\lambda_1+ \left( 4\,\lambda_1+4\,{\it u_5} \right) {{\it h_0}}^{3}
\nonumber\\
&&
+
\left( 4\,{\alpha_1}^{2}{\it u_1}+2\,{\alpha_3}^{2}{\it u_1}+4\,{\beta_1}^
{2}{\it u_3}+2\,{\beta_3}^{2}{\it u_3} \right) {\it h_0
}+4\,{\it u_4}\,\alpha_1\, \left( \alpha_1\,\beta_3+2\,\alpha_3\,\beta_1
\right)
\label{res5678}
\end{eqnarray}
where the first two equations describe the minimum of $V$ with respect to quark-antiquark components whereas the third and fourth equations describe this minimum with respect to four-quark composites.   The last equation minimizes the potential with respect to the scalar glueball field.  The  brackets with subscript zero represent evaluation of each derivative at VEV values
\begin{eqnarray}
\alpha_a &=& \langle S_a^a \rangle, \nonumber\\
\beta_a  &=& \langle {S'}_a^a \rangle, \nonumber \\
h_0  &=& \langle h \rangle.
\label{All_condensates}
\end{eqnarray}

The mass matrices for pions ($M^2_\pi$), kaons ($M^2_K$), system of $a_0$ scalars ($X^2_a$) and kappa system ($X^2_\kappa$) are:
\begin{eqnarray}
\left(M_\pi^2\right)_{11}
&=&
4\,{\it u_4}\,{\it h_0}\,\beta_3+2\,{\it u_1}\,{{\it h_0}}^{2}+4\,{\it u_2}
\,{\alpha_1}^{2}+2\,{\frac {{{\it h_0}}^{4}\lambda_2}{{\alpha_1}^{2}}}
\nonumber\\
&&+4\,
\left( -2+{
	\it \gamma_m} \right)
\left( {\it A_1}\,\alpha_1+1/2\,{\it A_3}\,\alpha_3 \right)
   \left( 2\,{\alpha_1}^{2}+{\alpha_3}^{2}
\right) ^{-{\it \gamma_m}/2}
\nonumber\\
\left(M_\pi^2\right)_{12}
&=&
4\,{\it u_4}\,{\it h_0}\,\alpha_3+{\frac {{  2\, {\it h_0}}^{4} \lambda_3  }{2\,\beta_1\,\alpha_1+\beta_3\,\alpha_3}}
\nonumber\\
\left(M_\pi^2\right)_{22}
&=&
2\,{\it u_3}\,{{\it h_0}}^{2}
\label{E_M2_pion_raw}
\end{eqnarray}

\begin{eqnarray}
\left(M_K^2\right)_{11}
&=&
		{1\over  {2\, {{\alpha_1}^{3}\alpha_3+\alpha_1\,{\alpha_3}^{3}}}}
	\left[
		4\,\alpha_1\, \left( -2+{\it \gamma_m} \right) \alpha_3\,
		\left( {\it A_1}\,\alpha_1+1/2\,{\it A_3}\,\alpha_3 \right)  \left( 2\,{
			\alpha_1}^{2}+{\alpha_3}^{2} \right) ^{1-{\it \gamma_m}/2}
     \right.
\nonumber \\
		&&		
		+4\, \left(
		{\alpha_1}^{2}+1/2\,{\alpha_3}^{2} \right)  \left( 2\,{\alpha_1}^{3}
		\alpha_3\,{\it u_2}-2\,{\alpha_1}^{2}{\alpha_3}^{2}{\it u_2}
\right.
\nonumber\\
&&	
\left.
\left.
	+\alpha_3\,
		\left( 2\,{\alpha_3}^{2}{\it u_2}+2\,{\it u_4}\,{\it h_0}\,\beta_1+{\it u_1
		}\,{{\it h_0}}^{2} \right) \alpha_1+{{\it h_0}}^{4}\lambda_2 \right)
		\right]
\nonumber\\
\left(M_K^2\right)_{12} &=&
		4\, \alpha_1\, u_4 \, h_0
		 + \frac{2\, h_0^4 \, \lambda_3}{2\,\beta_1\,\alpha_1+\beta_3\,\alpha_3}
\nonumber\\
\left(M_K^2\right)_{22} &=&
2\,{\it u_3}\,{{\it h_0}}^{2}
\label{E_M2_kaon_raw}
	\end{eqnarray}

\begin{eqnarray}
\left(X_a^2\right)_{11}
&=&
-4\,{\it u_4}\,{\it h_0}\,\beta_3+2\,{\it u_1}
\,{{\it h_0}}^{2}+12\,{\it u_2}\,{\alpha_1}^{2}-2\,{\frac {{{\it h_0}}^{4}
		\lambda_2}{{\alpha_1}^{2}}}
\nonumber\\
&&+4\, \left( -2+{\it \gamma_m} \right)
\left( {\it A_1}\,\alpha_1+1/2\,{\it A_3}\,\alpha_3 \right)  \left( 2\,{
	\alpha_1}^{2}+{\alpha_3}^{2} \right) ^{-{\it \gamma_m}/2}
\nonumber \\
\left(X_a^2\right)_{12}&=&
-4\,{\it u_4}\,{\it h_0}\,\alpha_3+{\frac {{  2 \,{\it h_0}}^{4} \lambda_3 }{2\,\beta_1\,\alpha_1+\beta_3\,\alpha_3}}
\nonumber\\
\left(X_a^2\right)_{22} &=&
2\,{\it u_3}\,{{\it h_0}}^{2}
\end{eqnarray}

\begin{eqnarray}
\left(X_\kappa^2\right)_{11}
&=&
-4\,{\it u_4}\,{\it h_0}\,\beta_1+2\,{\it u_1}\,{{\it h_0}}^{2}+4\,{\it u_2}
\, \left( {\alpha_1}^{2}+\alpha_1\,\alpha_3+{\alpha_3}^{2} \right) -2\,{
	\frac {{{\it h_0}}^{4}\lambda_2}{\alpha_1\,\alpha_3}}
\nonumber \\
&&+4\, \left( -2+{\it
	\gamma_m} \right)  \left( {\it A_1}\,\alpha_1+1/2\,{\it A_3}\,\alpha_3
\right)  \left( 2\,{\alpha_1}^{2}+{\alpha_3}^{2} \right) ^{-{\it
		\gamma_m}/2}
\nonumber\\
\left(X_\kappa^2\right)_{12} &=&
-4\,{\it u_4}\,{\it h_0}\,\alpha_1+{\frac {{ 2 {\it h_0}}^{4} \lambda_3 }{2\,\beta_1\,\alpha_1+\beta_3\,\alpha_3}}
\nonumber\\
\left(X_\kappa^2\right)_{22} &=& 2\,{\it u_3}\,{{\it h_0}}^{2}
\end{eqnarray}

Note that these mass matrices are subject to the vacuum conditions expressed by Eqs. (\ref{res5678}).   Once these vacuum conditions are invoked some of the unknown parameters can be determined in terms of the rest of the parameters.  For example,  we can use the five equations given in Eqs. (\ref{res5678}) to solve for $u_1\cdots u_5$ (these solutions  are rather bulky and we do not give them here).   Upon substitution of these solutions  back into the pion mass matrix (\ref{E_M2_pion_raw}) we can compute the determinant

\begin{eqnarray}
\det\left(M^2_\pi\right) = 
-4\,{\frac { \left( 2\,\alpha_1^{2}+{\alpha_3}^{2} \right) ^{1-{\gamma_m}/2}\,{{h_0}}^{4}\lambda_3\, \left( {\alpha_1}^{2}-
{\alpha_3}^{2} \right) A_1 }{\alpha_1\, \left( 2\,{\alpha_1}^{2}{\beta_1}^{2}-
\alpha_1\,\alpha_3\,\beta_1\,\beta_3-{\alpha_3}^{2}{\beta_3}^{2} \right) }}
\label{E_detM2pi_subs}
\end{eqnarray}
which is, as expected, proportional to the quark masses (here proportional to $A_1$) and vanishes when there is no explicit symmetry breaking.   Since $\det\left(M^2_\pi\right) = m_\pi^2 {m'}_\pi^2$, Eq. (\ref{E_detM2pi_subs}) results in a massless pion when $A_1\rightarrow 0$.   

Similarly, using the kaon mass matrix given above, together with vacuum conditions  (\ref{res5678}), we can show

\begin{eqnarray}
\det\left(M^2_K\right) &=&
-8\,\lambda_3\, 
\left( -\alpha_3+\alpha_1 \right) h_0^4 \left( 2\,{\alpha_1}^{2}+{ \alpha_3}^{2} \right)^{1-{\gamma_m}/2}
\Bigg[ 
     -    \left( {A_1}\, \alpha_1+   { A_3\over 2}\,\alpha_3 \right)  
          \left( -2+{\gamma_m} \right)  
          \left( \alpha_1+\alpha_3 \right)  
          \nonumber\\
&&
+  
         \left( {A_1}\,{\it \gamma_m}-{ A_1}+{A_3} \right) 
     {\alpha_1}^{2}+ 
     \left( -2+{ \gamma_m} \right)  
     \left( {A_1}+{A_3\over 2} \right) 
     \alpha_3\,\alpha_1
+{{\alpha_3}^{2}\over 2} 
     \left( {A_3}\,{\gamma_m}+{ A_1}-{ A_3} \right)   
\Bigg]
\nonumber\\
&&
\Bigg/
\Bigg[
\left( 2\,{\alpha_1}^{2}+{\alpha_3}^{2} \right)  
\left( 2\,\beta_1\,\alpha_1+\beta_3\,\alpha3 \right)  
\left( \beta_1\,\alpha_1-\beta_3\,\alpha_3 \right)
\Bigg]
\end{eqnarray}
which clearly shows that 
\begin{equation}
\lim_{A1,A_3\rightarrow 0}\det\left(M^2_K\right) \rightarrow 0
\end{equation}
and since $\det\left(M^2_K\right) = m_K^2 {m'}_K^2$, in the absence of quark masses $m_K$ vanishes.

The mass matrices for $f_0$ and $\eta$ systems (each a $5\times 5$ matrix) are more involved and given  in Appendix A.  
The determinant of eta system is also explicitly given in Appendix A and shown to be proportional to quark masses (similar to the above cases for pion and kaon systems).

Note that in the absence of the glueballs,  the mass matrices obtained in this section as well as the minimum equations should agree with those given in \cite{Jora5}.  In order to check this, we first make the substitution for $u_1 \cdots u_4$ and $\gamma_m$ in terms of $c_2, c_4^a, d_2$ and $e_3^a$ defined in \cite{Jora5}:
\begin{eqnarray}
u_1\rightarrow u_1^0 &=& -{c_2 \over {h_0^2}} \nonumber\\
u_2\rightarrow u_2^0 &=& c_4^a \nonumber \\
u_3\rightarrow u_3^0 &=& {d_2\over {h_0^2}} \nonumber\\
u_4\rightarrow u_4^0 &=& e_3^a \over h_0\nonumber\\
\gamma_m \rightarrow \gamma_m^0 &=& 2
\label{mmp_subs}
\end{eqnarray}
and then take the limit of $h_0\rightarrow 0$:
\begin{eqnarray}
\left( X^2_a \right)^{(0)} &=& \lim_{h_0\rightarrow 0}\left[ \left( X^2_a \right)\Big|_{u_i = u^0_i, \gamma_m=
	\gamma_m^0}\right]\nonumber\\
\left( X^2_\kappa \right)^{(0)} &=& \lim_{h_0\rightarrow 0}\left[ \left( X^2_\kappa \right)\Big|_{u_i = u^0_i, \gamma_m=
	\gamma_m^0}\right]\nonumber\\
\left( X^2_0 \right)^{(0)} &=& \lim_{h_0\rightarrow 0}\left[ \left( X^2_0 \right)\Big|_{u_i = u^0_i, \gamma_m=
	\gamma_m^0}\right]\nonumber\\
\left( M^2_\pi \right)^{(0)} &=& \lim_{h_0\rightarrow 0}\left[ \left( M^2_\pi \right)\Big|_{u_i = u^0_i, \gamma_m=
	\gamma_m^0}\right] \nonumber \\
\left( M^2_K \right)^{(0)} &=& \lim_{h_0\rightarrow 0}\left[ \left( M^2_K \right)\Big|_{u_i = u^0_i, \gamma_m=
	\gamma_m^0}\right]
\end{eqnarray}
where $i = 1 \cdots 4$.   We find that the expected limits are upheld.

The case of $M^2_0$ this is more complicated. There is a clear limit in which the glueball part in the Lagrangian in  Eq. (\ref{inlgr567}) can lead in first order to the axial anomaly term in Eq. (\ref{SpecLag}). This is achieved when the bare glueball mass term $u_6h^2g^2$ or more exactly $u_6$ is very large such that the kinetic term becomes negligible and the glueball field can be integrated out. Therefore:
\begin{eqnarray}
\frac{\partial V}{\partial g}&=&
u_6h^2g -i\frac{h^3}{12}\left[ \gamma_1\ln \left(\frac{\det M}{\det M^{\dagger}}\right)+\gamma_2\ln\left(\frac{{\rm Tr}(MM^{\prime\dagger})}{{\rm Tr}(M^{\prime}M^{\dagger})}\right)\right]=0,
\label{gluebint775}
\end{eqnarray}
and solve for $g$ to obtain:
\begin{eqnarray}
g=i\frac{h}{24u_6}\left[ \gamma_1\ln \left(\frac{\det M}{\det M^{\dagger}}\right)+\gamma_2\ln\left(\frac{{\rm Tr}(MM^{\prime\dagger})}{{\rm Tr}(M^{\prime}M^{\dagger})}\right)\right].
\label{solg5665}
\end{eqnarray}
which upon substitution into the pseudoscalar glueball piece of the Lagrangian (\ref{solg5665}) leads to the following term:
\begin{eqnarray}
V_{\rm A}^{\rm eff}=\frac{h^4}{288u_6}\left[ \gamma_1\ln \left(\frac{\det M}{\det M^{\dagger}}\right)+\gamma_2\ln\left(\frac{{\rm Tr}(MM^{\prime\dagger})}{{\rm Tr}(M^{\prime}M^{\dagger})}\right)\right]^2,
\label{res639998}
\end{eqnarray}
In first order this leads to identification of $c_3$ in Eq. (\ref{SpecLag}) with:
\begin{eqnarray}
u_6 \rightarrow u_6^0 = \frac{h_0^4}{288\, c_3}.
\label{ident7756}
\end{eqnarray}
Therefore, we expect
\begin{eqnarray}
\left( M^2_\eta \right)_{\alpha \beta}^{(0)} &=& \lim_{h_0\rightarrow 0}\left[ \left( M^2_\eta \right)_{\alpha \beta}\Big|_{u_i = u^0_i, \gamma_m=
	\gamma_m^0}\right] +
\lim_{u_6\rightarrow u_6^0}
\left[
\left\langle {{\partial^2 V_{\rm A}^{\rm eff}}\over {\partial \eta_\alpha \partial \eta_\beta}} \right\rangle_0
\right]
\end{eqnarray}
where $i, \alpha, \beta = 1 \cdots 4$.    We have verified that this equation is satisfied as well.

\section{Decoupling limit}
We consider a limiting case where the glueball fields are decoupled from the quark mesons and affect the system only through the vacuum.    This limit is important because (a) it allows probing the pure glueball mass from the stability of vacuum, and (b) it defines  a basic starting point for tackling  the complicated system of mass matrices and minimum equations (derived in previous section) and makes it possible to study the mass spectrum and the interaction vertices as well as the spectroscopy of the physical scalar and pseudoscalar states in which the formation of the quark and gluball components are assembled step by step upon the properties of the vacuum.  Particularly, it is well known that disentangling the two-quark,  four-quark and glueball  building blocks of isosinglet states (particularly scalars) is a non-trivial undertaking,  and in this approach,  we begin with a careful study of the vacuum containing quark-antiquarks, four-quarks and non-interacting glueballs.

For comparison we refer to the work of \cite{Jora41,Jora42,Jora43}  in SU(3) limit.    In these works the effect of pseudoscalar glueball was fully taken into account (in which while the pseudoscalar glueball is integrated out, the U(1)$_{\rm A}$ is exactly saturated), but in \cite{Jora41,Jora42,Jora43}  no scalar glueballs were present.   In the decoupling limit of the present work,  the situation is rather reversed: the pseudoscalar glueball is completely decoupled and only non-interacting scalar glueballs are considered.   We find that, expectedly, the eta masses are not physical in this limit (demonstrating the well-known fact about the importance of the axial anomaly), whereas the scalar isosinglet masses (and their quark substructure) are similar to those found in the three references just cited which shows that in oder to make physical predictions,  having scalar glueball in the vacuum is not sufficient and the glueball interactions with quark mesons should be turned on.

For the scalars the decoupling conditions for the choice $\gamma_m=2$ are straightforward:
\begin{eqnarray}
&&[X_0^2]_{45}=4\alpha_1^2u_4+4u_3h_0\beta_3+\frac{8h_0^3\lambda_3\alpha_3}{2\alpha_1\beta_1+\alpha_3\beta_3}=0
\nonumber\\
&&[X_0^2]_{35}=4\sqrt{2}\alpha_1\alpha_3u_4+4\sqrt{2}u_3h_0\beta_1+\frac{8\sqrt{2}h_0^3\lambda_3\alpha_1}{2\alpha_1\beta_1+\alpha_3\beta_3}=0
\nonumber\\
&&[X_0^2]_{25}=8u_4\alpha_1\beta_1+4u_1h_0\alpha_3+\frac{8h_0^3\lambda_2}{\alpha_3}+\frac{8h_0^3\lambda_3\beta_3}{2\alpha_1\beta_1+\alpha_3\beta_3}=0
\nonumber\\
&&[X_0^2]_{15}=4\sqrt{2}u_4[\beta_1\alpha_3+\alpha_1\beta_3]+4\sqrt{2}u_1h_0\alpha_1+\frac{8\sqrt{2}h_0^3\lambda_2}{\alpha_1}+\frac{8\sqrt{2}h_0^3\lambda_3\beta_1}{2\alpha_1\beta_1+\alpha_3\beta_3}.
\label{decoupl865748}
\end{eqnarray}
The two minimum equations of interest refer to $M'$ components and are:
\begin{eqnarray}
&&4\alpha_1^3\beta_1u_4+2\alpha_1^2\alpha_3\beta_3u_4+2\alpha_1\beta_1\beta_3h_0u_3+\alpha_3\beta_3^2h_0u_3+\alpha_3h_0^3\lambda_3=0
\nonumber\\
&&4\alpha_1^2\alpha_3\beta_1u_4+2\alpha_3^2\beta_3\alpha_1u_4+2\beta_1^2\alpha_1h_0u_3+h_0^3\lambda_3\alpha_1+\alpha_3\beta_1\beta_3h_0u_3=0
\label{mineq555243}
\end{eqnarray}
These equations may be put in more amenable form:
\begin{eqnarray}
&&2\alpha_1^2u_4+\beta_3h_0u_3+\frac{\alpha_3h_0^3\lambda_3}{2\alpha_1\beta_1+\alpha_3\beta_3}=0
\nonumber\\
&&2\alpha_1\alpha_3u_4+\beta_1h_0u_3+\frac{\alpha_1h_0^3\lambda_3}{2\alpha_1\beta_1+\alpha_3\beta_3}=0.
\label{mineq628889}
\end{eqnarray}

Subtracting first equation in Eq. (\ref{decoupl865748}) from the second equation in Eq. (\ref{mineq628889}) we obtain:
\begin{eqnarray}
\lambda_3=\frac{\alpha_1^2u_4(2\alpha_1\beta_1+\alpha_3\beta_3)}{h_0^3\alpha_3}.
\label{lambda3_sol}
\end{eqnarray}
Then the system of four equations reduces to:
\begin{eqnarray}
&&3\alpha_1^2u_4+u_3h_0\beta_3=0
\nonumber\\
&&2\alpha_1\alpha_3u_4+\frac{\alpha_1^3}{\alpha_3}u_4+\beta_1h_0u_3=0
\nonumber\\
&&\alpha_1\alpha_3u_4+\frac{2\alpha_1^3}{\alpha_3}u_4+\beta_1h_0u_3=0
\label{res674665}
\end{eqnarray}
Solving this system leads to:
\begin{eqnarray}
&&\alpha_3=\alpha_1=\alpha
\nonumber\\
&&\beta_3=\beta_1=\beta
\nonumber\\
&&u_3=-\frac{3\alpha^2 u_4}{\beta h_0}.
\label{u3_sol}
\end{eqnarray}
We thus arrive at the SU(3)$_{\rm V}$ limit. From the first two minimum equations  we determine $A_1=A_3=A$. We further solve the last two equations (which are identical in the SU(3)$_{\rm V}$ limit) in Eq. (\ref{decoupl865748}) to determine:
\begin{eqnarray}
\lambda_2=-\frac{\alpha^2(h_0u_1+4\beta u_4)}{2h_0^3}.
\label{lambda2_sol}
\end{eqnarray}
In the SU(3)$_{\rm V}$ limit the mass matrices are organized in terms of octet-singlet bases in which
the mass matrices for  scalars ($Y^2$) and pseudoscalars ($N^2$) are related to the isosinglet scalar and pseudoscalar mass matrices by
\begin{eqnarray}
Y^2 &=& {\widetilde T} \, X_0^2\, T  \nonumber\\
N^2 &=& {\widetilde T} \, M_0^2\,   T
\label{E_NY_definitions}
\end{eqnarray}
where $T$ is the transformation matrix between strange-nonstrange (SNS) basis and the octet-singlet (OS) basis which its explicit form is not unique and obviously depends on the specific way that the basis vectors are defined and organized.   Our definitions and notations for these bases (and our preference for their organization) are as follows:  
Our notations for  physical and bare (in both SNS and OS bases) scalar and pseudoscalar bases are
\begin{equation}
F_{_{\rm phy.}} =
\left[
\begin{array}{cc}
f_1\\
f_2\\
f_3\\
f_4\\
f_5
\end{array}
\right],
\hskip .75cm
F_{_{\rm SNS}}=
\left[
\begin{array}{cc}
f_a\\
f_b\\
f_c\\
f_d\\
h
\end{array}
\right],
\hskip .75cm
F_{_{\rm OS}}=
\left[
\begin{array}{cc}
f_8\\
f'_8\\
f_0\\
f'_0\\
h
\end{array}
\right],
\hskip .75cm
\eta_{_{\rm phy.}} =
\left[
\begin{array}{cc}
\eta_1\\
\eta_2\\
\eta_3\\
\eta_4\\
\eta_5
\end{array}
\right],
\hskip .75cm
\eta_{_{\rm SNS}}=
\left[
\begin{array}{cc}
\eta_a\\
\eta_b\\
\eta_c\\
\eta_d\\
g
\end{array}
\right],
\hskip .75cm
\eta_{_{\rm OS}}=
\left[
\begin{array}{cc}
\eta_8\\
\eta'_8\\
\eta_0\\
\eta'_0\\
g
\end{array}
\right],
\label{F_eta_OS}
\end{equation}
where $f_1 \cdots f_5$ and $\eta_1 \cdots \eta_5$ are respectively the five lowest physical isosinglet scalars and pseudoscalars and
\begin{eqnarray}
f_a&=&\frac{S^1_1+S^2_2}{\sqrt{2}} \hskip .7cm
\propto \hskip .5cm n{\bar n},
\nonumber  \\
f_b&=&S^3_3 \hskip 1.6 cm \propto \hskip .5cm s{\bar s},
\nonumber    \\
f_c&=&  \frac{S'^1_1+S'^2_2}{\sqrt{2}}
\hskip .5 cm \propto \hskip .5cm ns{\bar n}{\bar s},
\nonumber   \\
f_d&=& S'^3_3
\hskip 1.5 cm \propto \hskip .5cm nn{\bar n}{\bar n},
\label{f_basis}
\end{eqnarray}
are the bare quark-antiquark and four-quark components in the SNS scalar basis.  In the OS basis
\begin{eqnarray}
f_8 &=& \frac{S^1_1 + S^2_2 - 2 S_3^3}{\sqrt{6}},
\nonumber  \\
f'_8 &=& \frac{S'^1_1 + S'^2_2 - 2S'^3_3}{\sqrt{6}},
\nonumber  \\
f_0 &=& \frac{S^1_1 + S^2_2 + S_3^3}{\sqrt{3}},
\nonumber  \\
f'_0 &=& \frac{S'^1_1 + S'^2_2  + S'^3_3}{\sqrt{3}},
\label{f_OS_basis}
\end{eqnarray}
where the scalar components $f_8$ and $f_0$ are of quark-antiquark type whereas the $f'_8$ and $f'_0$ have  four-quark substructure.  Similarly,  
\begin{eqnarray}
\eta_a&=&\frac{\phi^1_1+\phi^2_2}{\sqrt{2}} \hskip .7cm
\propto \hskip .5cm n{\bar n},
\nonumber  \\
\eta_b&=&\phi^3_3 \hskip 1.6 cm \propto \hskip .5cm s{\bar s},
\nonumber    \\
\eta_c&=&  \frac{\phi'^1_1+\phi'^2_2}{\sqrt{2}}
\hskip .5 cm \propto \hskip .5cm ns{\bar n}{\bar s},
\nonumber   \\
\eta_d&=& \phi'^3_3
\hskip 1.5 cm \propto \hskip .5cm nn{\bar n}{\bar n}.
\label{eta_basis}
\end{eqnarray}
are the bare quark-antiquark and four-quark pseudoscalar components in the SNS basis.  In the OS basis
\begin{eqnarray}
\phi_8 &=& \frac{\phi^1_1 + \phi^2_2 - 2 \phi_3^3}{\sqrt{6}},
\nonumber  \\
\phi'_8 &=& \frac{\phi'^1_1 + \phi'^2_2 - 2\phi'^3_3}{\sqrt{6}},
\nonumber  \\
\phi_0 &=& \frac{\phi^1_1 + \phi^2_2 + \phi_3^3}{\sqrt{3}},
\nonumber  \\
\phi'_0 &=& \frac{\phi'^1_1 + \phi'^2_2  + \phi'^3_3}{\sqrt{3}},
\label{eta_OS_basis}
\end{eqnarray}
where the pseudoscalar components $\phi_8$ and $\phi_0$ are of quark-antiquark type whereas the $\phi'_8$ and $\phi'_0$ have  four-quark substructure.
The SNS and OS bases are related by transformation matrix $T$
\begin{eqnarray}
F_{_{\rm SNS}} &=& T F_{_{\rm OS}}, \nonumber\\
\eta_{_{\rm SNS}} &=& T \eta_{_{\rm OS}},
\end{eqnarray}
where, in our setup, matrix $T$ has the explicit form
\begin{equation}
T =
\left[
\begin{array}{ccccc}
1\over \sqrt {3} &           0 &  \sqrt{2\over3}&          0& 0 \\
- \sqrt{2\over3}&           0& 1\over \sqrt {3}&         0& 0 \\
0 &   1\over \sqrt {3} &         0&  \sqrt{2\over3} & 0 \\
0 & - \sqrt{2\over3} &         0& 1\over \sqrt {3} & 0 \\
0 &           0 &         0 &         0&  1
\end{array}
\right].
\end{equation}

The mass matrices $Y^2$ and $N^2$ defined in Eq. (\ref{E_NY_definitions}) have a block diagonal octet singlet structure
\begin{eqnarray}
\left[Y^2 \right]_{5\times 5 } &=&
\left[
\begin{array}{cccc}
&                                     &                                       &   \\
&    \left[Y_8^2\right]_{2 \times 2}  &                                       &   \\
&                                     &                                       &   \\
&                                     &  \left[Y_0^2\right]_{3 \times 3}      &   \\
&                                     &                                       &   \\
\end{array}
\right],
\nonumber \\
\left[N^2 \right]_{5\times 5 } &=&
\left[
\begin{array}{cccc}
&                                     &                                   &   \\
&    \left[N_8^2\right]_{2 \times 2}  &                                   &   \\
&                                     &                                   &   \\
&                                     &  \left[N_0^2\right]_{3 \times 3}  &   \\
&                                     &                                   &
\end{array}
\right].
\end{eqnarray}
The octet physical states
\begin{equation}
\Psi_{8^+}  =
\left[
\begin{array}{cc}
\psi_{8^+}^{(1)}\\
\psi_{8^+}^{(2)}
\end{array}
\right],
\hskip .75cm
\Psi_{8^-}  =
\left[
\begin{array}{cc}
\psi_{8^-}^{(1)}\\
\psi_{8^-}^{(2)}
\end{array}
\right],
\end{equation}
diagonalize $\left[Y_8^2\right]$ and $\left[N_8^2\right]$ respectively and are related to the octet ``bare'' states
\begin{equation}
B_{8^+}=
\left[
\begin{array}{c}
f_8\\
f'_8
\end{array}
\right],
\hskip .75cm
B_{8^-}=
\left[
\begin{array}{c}
\eta_8\\
\eta'_8
\end{array}
\right],
\label{F_eta_OS}
\end{equation}
by
\begin{eqnarray}
\Psi_{8^+} &=&
\left[{K_{8^+}}\right]^{-1}
B_{8^+},
\nonumber\\
\Psi_{8^-} &=&
\left[{K_{8^-}}\right]^{-1}
B_{8^-},
\end{eqnarray}
therefore
\begin{eqnarray}
{\widetilde \Psi}_{8^+}
 \left[ Y_8^2\right]_{\rm diag}  \Psi_{8^+} & = &
{\widetilde  B_{8^+}}  \left[Y_8^2\right] B_{8^+}, \nonumber \\
{\widetilde \Psi}_{8^-}
 \left[ N_8^2\right]_{\rm diag}  \Psi_{8^-} & = &
 {\widetilde  B_{8^-}}  \left[N_8^2\right] B_{8^-}.
\end{eqnarray}

In the present decoupling limit, $Y_0^2$ and $N_0^2$ themselves become block diagonal as well:
\begin{eqnarray}
Y_0^2 &=&
\left[
\begin{array}{cccc}
&                                     &                                   &   \\
&    \left[{\widehat Y}_0^2\right]_{2 \times 2}  &                            &   \\
&                                     &                                   &   \\
&                                     &                m_h^2              &   \\
&                                     &                                   &
\end{array}
\right]
\nonumber \\
N_0^2 &=&
\left[
\begin{array}{cccc}
&                                     &                                   &   \\
&    \left[{\widehat N}_0^2\right]_{2 \times 2}  &                                   &   \\
&                                     &                                   &   \\
&                                     &              m_g^2                &   \\
&                                     &                                   &
\end{array}
\right]
\end{eqnarray}
In this case, the physical singlet states
\begin{equation}
{\widehat \Psi}_{0^+}  =
\left[
\begin{array}{cc}
{\widehat\psi}_{0^+}^{(1)}\\
{\widehat\psi}_{0^+}^{(2)}
\end{array}
\right],
\hskip .75cm
{\widehat \Psi}_{0^-}  =
\left[
\begin{array}{cc}
{\widehat\psi}_{0^-}^{(1)}\\
{\widehat\psi}_{0^-}^{(2)}
\end{array}
\right],
\end{equation}
are related to the ``bare'' singlet states
\begin{equation}
{\widehat B}_{0^+}=
\left[
\begin{array}{c}
{\widehat f}_0\\
{\widehat f}'_0
\end{array}
\right],
\hskip .75cm
{\widehat B}_{0^-}=
\left[
\begin{array}{c}
{\widehat \eta}_0\\
{\widehat \eta}'_0
\end{array}
\right],
\label{F_eta_OS}
\end{equation}
by
\begin{eqnarray}
{\widehat\Psi}_{0^+} &=&
\left[{{\widehat K}_{0^+}}\right]^{-1}
{\widehat  B}_{0^+}
\nonumber\\
{\widehat \Psi}_{0^-} &=&
\left[{{\widehat K}_{0^-}}\right]^{-1}
{\widehat B}_{0^-}
\end{eqnarray}
which means
\begin{eqnarray}
{\widetilde {\widehat \Psi}}_{0^+}
\left[ {\widehat Y}_0^2\right]_{\rm diag}  {\widehat \Psi}_{0^+} & = &
{\widetilde {\widehat  B}}_{0^+}  \left[{\widehat Y}_0^2\right] {\widehat B}_{0^+}, \nonumber \\
{\widetilde {\widehat \Psi}}_{0^-}
\left[ {\widehat N}_0^2\right]_{\rm diag}  {\widehat \Psi}_{0^-} & = &
{\widetilde {\widehat  B}}_{0^-}  \left[{\widehat N}_0^2\right] {\widehat B}_{0^-},
\end{eqnarray}

The mass matrices $Y^2_8$, $Y^2_0$, $N^2_8$ and $N^2_0$ in the SU(3) limit are
\begin{eqnarray}
\left(Y^2_8\right)_{11} &=&
{1\over \alpha^2}
\left(12\,u_2\,{\alpha}^{4}-4\, u_4 \,  h_0\,\beta\,{
	\alpha}^{2}+2\, u_1\,{ h_0}^{2}{\alpha}^{2}-2\,{ h_0}^{4}
\lambda_2\right)
\nonumber\\
\left(Y^2_8\right)_{12} &=& 		
{{2h_0}\over {3\alpha\beta}}
\left( -6\,{\alpha}^{2}\beta\,  u_4 +{h_0}^{3}\lambda_3 \right)
\nonumber\\
\left(Y^2_8\right)_{22} &=& 2 u_3 h_0^2		
\end{eqnarray}

\begin{eqnarray}
\left(Y^2_0\right)_{11} &=&
{1 \over {3\alpha^2}}\left[
\left( -6\,\lambda_2-2\,\lambda_3 \right) h_0^{4}+6
\, u_1\, h_0^2{\alpha}^{2}+24\, u_4 \, h_0\,\beta\,
{\alpha}^{2}+36\, u_2\,{\alpha}^{4}
\right]
\nonumber\\
\left(Y^2_0\right)_{12} &=&
8\,u_4\, h0\,\alpha
\nonumber\\
\left(Y^2_0\right)_{13} &=&
{4\over {\sqrt{3}\alpha}}\,
\left( 6\,{\alpha}^{2}\beta\,{\it u_4}+3\,{\it
	u_1}\,{\it h_0}\,{\alpha}^{2}+6\,{h_0}^{3}\lambda_2+2\,{h_0}^{3
}\lambda_3 \right)
\nonumber\\
\left(Y^2_0\right)_{22} &=&
-{{2h_0^2}\over {3 \beta^2}} \,
\left( -3\, u_3\, \beta^{2}+ h_0^2\lambda_3 \right)
\nonumber\\		
\left(Y^2_0\right)_{23}		&=&	
{{4\sqrt{3}}\over {3\beta}}
\left( 3\,{\alpha}^{2}\beta\,{\it u_4}+3\, u_3\, h_0\,{\beta}^{2}+2\, h_0^{3}\lambda_3 \right)
\nonumber\\
\left(Y^2_0\right)_{33}		&=&	
24\,\ln  \left( {\frac {{\alpha}^{3}}{{\Lambda}^{3}}} \right) h_0^2 \lambda_2
+12\,\ln  \left( {\frac {h_0^4}{{\Lambda}^{4}}}
\right) {h_0}^{2}\lambda_1+24\,\ln  \left( {\frac {\alpha\,\beta
	}{{\Lambda}^{2}}} \right) {h_0}^{2}\lambda_3+24\,\ln  \left( 3
\right) { h_0}^{2}\lambda_3 + \left( 28\,\lambda_1 + 12\,{\it u_5}
\right) {h_0}^{2}
\nonumber\\
&&
+6\, u_1 \,{\alpha}^{2}+6\, u_3 \,{\beta}
^{2}
\end{eqnarray}

\begin{eqnarray}
\left(N^2_8\right)_{11} &=&
{1\over {\alpha^2}}
\left(
4\, u_2 \, \alpha^4 + 4\,  u_4 \, h_0 \, \beta \, \alpha^2 + 2\,  u_1 \, h_0^2 \alpha^2 + 2\, h_0^4
\lambda_2
\right)
\nonumber\\
\left(N^2_8\right)_{12} &=&
{{2 h_0}\over {3 \alpha \beta}}
\left( 6\, \alpha^2 \beta\,  u_4 +  h_0^3 \lambda_3 \right)
\nonumber\\
\left(N^2_8\right)_{22} &=&
2\, u_3\,  h_0^2
\end{eqnarray}

\begin{eqnarray}
\left(N^2_0\right)_{11} &=&
{1\over {3\alpha^2}}
\left[
\left(
6\,\lambda_2 + 2\,\lambda_3 \right) h_0^4 + 6 \,  u_1 \, h_0^2 \alpha^2 - 24\, u_4 \,  h_0 \, \beta\,
\alpha^2 + 12\, u_2 \, \alpha^4
\right]
\nonumber\\
\left(N^2_0\right)_{12} &=& -8\, u_4 \, h_0\, \alpha
\nonumber\\
\left(N^2_0\right)_{13} &=&
{{\sqrt{3}h_0^3}\over {18\alpha}}
\left( 2\,\gamma_1 + 1 \right)
\nonumber\\
\left(N^2_0\right)_{22} &=&
{{2h_0^2}\over {3\beta^2}}\,
\left( 3\,u_3\,\beta^2 + h_0^2\lambda_3 \right)
\nonumber\\
\left(N^2_0\right)_{23} &=&
{{\sqrt {3} h_0^3}\over {18\beta}}\,  \left( -1+\gamma_1 \right)
\nonumber\\
\left(N^2_0\right)_{33} &=& 2\, u_6\, h_0^2
\nonumber\\
\end{eqnarray}

In the decoupling limit $\left(Y^2_0\right)_{13}=\left(Y^2_0\right)_{23}=\left(N^2_0\right)_{13}=\left(N^2_0\right)_{23}=0$.

The trace of the mass matrices for the scalar and pseudoscalar octets are the sum of the physical masses,
\begin{eqnarray}
{[N_8^2]}_{11}+{[N_8^2]}_{22} &=& m_{8^-}^2 + {m'}_{8^-}^2
\nonumber\\
{[Y_8^2]}_{11}+{[Y_8^2]}_{22} &=& m_{8^+}^2 + {m'}_{8^+}^2,
\label{masstr64553}
\end{eqnarray}
which can be used to calculate $u_1$ and $u_4$ in terms of $\alpha_1$, $\beta_1$ and $h$:
\begin{eqnarray}
u_1 &=& -\frac{m_{8^-}^2 + m_{8^-}^{\prime 2} + 8\alpha^2 u_2 - m_{8^+}^2 - {m'}_{8^+}^2} {2 h_0^2}.
\nonumber\\
u_4 &=& -\frac{\beta \left( 3 m_{8^-}^2 + 3 {m'}_{8^-}^2 -  m_{8^+}^2 -  {m'}_{8^+}^2 \right)} {12\alpha^2 h_0}.
\label{u1u4_sol}
\end{eqnarray}
We have two more relations for the determinants of the octet scalar and pseudoscalar mass matrices. In order to simplify them we will  subtract and add the two determinants which leads to:
\begin{eqnarray}
\left(  m_{8^-}^2 + {m'}_{8^-}^2 - m_{8^+}^2 - {m'}_{8^+}^2 \right)
\left[N_8^2\right]_{22} - \left( \left[N_8^2\right]_{12}^2 - \left[Y_8^2\right]_{12}^2 \right) &=& m_{8^-}^2 m_{8^-}^{\prime 2}  - m_{8^+}^2 m_{8^+}^{\prime 2}
\nonumber\\
\left[m_{8^-}^2 + {m'}_{8^-}^2 + (m_{8^+}^2 + {m'}_{8^+}^2 )\right] \left[N_{8^-}^2\right]_{22}
- 2\left(
          \left[N_{8^-}^2\right]_{22}^2 - \left[N_{8^-}^2\right]_{12}^2 - \left[Y_8^2\right]_{12}^2
    \right) &=&
m_{8^-}^2 {m'}_{8^-}^2 + m_{8^+}^2 m_{8^+}^2,
\label{resuuyyyy677}
\end{eqnarray}
From the first equation we determine $\beta$ in terms of $\alpha$:
\begin{eqnarray}
{\beta\over \alpha} &=& -{3\over 2} \,
\Bigg[
3\,\left( m_{8^-}^4  + {m'}_{8^-}^4  \right)  +  m_{8^+}^4 +  {m'}_{8^+}^4 +
			4 \left( m_{8^-}^2  -  m_{8^+}^2 -   {m'}_{8^+}^2 \right)  {m'}_{8^-}^2			
\nonumber\\
&&
\hskip .75cm			+ 4 \left(  m_{8^+}^2 -   m_{8^-}^2 \right)  {m'}_{8^+}^2
			-4\,  m_{8^+}^2  m_{8^-}^2
\Bigg]^{1/2}	
	\left(  m_{8^+}^2 +   {m'}_{8^+}^2 - 3\,  m_{8^-}^2 - 3\,  {m'}_{8^-}^2\right)	
\label{beta_over_alpha}
\end{eqnarray}

 The second equation is then independent of any parameter and is a constraint applied to the physical mass of the heavy pseudoscalar octet:
 \begin{eqnarray}
  {m'}_{8^-} &=&
  {2\over 3}\, \left( m_{8^+}^2 + {m'}_{8^+}^2\right)  - {38 \over 39} \, m_{8^-}^2
 \nonumber \\
 && \hskip .75cm
 \pm
 {1 \over 39}
 \Bigg[
 169\,  m_{8^+}^4  +
 \left(
       -364\,  {m'}_{8^+}^2
        + 52\, m_{8^-}^2
 \right) m_{8^+}^2
+ 169\, {m'}_{8^+}^4
+ 52\,
{m'}_{8^+}^2  {m}_{8^-}^2
- 77\, m_{8^-}^4
\Bigg]^{1/2}
  \label{mp_8m}
 \end{eqnarray}

Using the octect decay constant and Eq. (\ref{beta_over_alpha}), we determine $\alpha$
\begin{equation}
\alpha =
{f_{8^-} \over
 {2\left( \cos \theta_{8^-} - {\beta\over \alpha}  \sin \theta_{8^-} \right) } }
\end{equation}
where
\begin{eqnarray}
\cos \theta_{8^-}=\Bigg(
\frac{	2u_3h_0^2  - m_{8^-}^2}
     {{m'}_{8^-}^2 - m_{8^-}^2}
     \Bigg)^{1/2} =
\sqrt {
	{\frac
		{m_{8^-}^2   + 3\, {m'}_{8^-}^2 - m_{8^+}^2 - {m'}_{8^+}^2}
		{2 \left({m'}_{8^-}^2  - m_{8^-}^2\right) }
	}
},
\label{res5536677}
\end{eqnarray}
and this, together with (\ref{beta_over_alpha}),  determines $\beta$.   Upon substitution  of $\alpha$ and $\beta$ into (\ref{u1u4_sol}) $u_4$ is determined as a function of $h_0$ [i.e. $u_4(h_0)$].  When $u_4(h_0)$ is substituted in (\ref{lambda3_sol}), results in  $\lambda_2(h_0)$.
Using a first order estimate stemming from the trace anomaly results (namely $\lambda_1=\frac{11}{36}$), together with $\lambda_3(h_0)$,  result in $\lambda_2(h_0)$   which in turn, when substituted in Eq. (\ref{lambda2_sol}), determines $u_1(h_0)$.   Finally, when $u_1(h_0)$ is substituted in the first equation of (\ref{u1u4_sol}) determines  $u_2(h_0)$.   Therefore, all parameters are determined in terms of $h_0$.

For numerical analysis we examine the following inputs for the octets (and then make variations to study the sensitivity of the results):
\begin{eqnarray}
m_{8^-} &=& 137 \,\,{\rm MeV}
\nonumber\\
f_{8^-} &=& 131 \,\,{\rm MeV}
\nonumber\\
m_{8^+} &=& 980 \,\,{\rm MeV}
\nonumber\\
{m'}_{8^+} &=& 1474\,\,{\rm MeV}
\label{res7665774}
\end{eqnarray}

With these inputs, Eq. (\ref{mp_8m}) gives ${m'}_{8^-}$ = 1308 MeV which is near the central value of the experimental data on $m[\pi(1300)]$ = 1.2-1.4 GeV \cite{PDG}.   In Table \ref{T_param}  our results for the model parameters are given for two values of $h_0$.   These model parameters then allow predictions for SU(3) singlet masses (Table \ref{T_SU3_singlet_masses}) and rotation matrices (Table \ref{T_Rotation_matrices}), both being independent of $h_0$.     This is of course expected in the decoupling limit in which the properties of scalar and pseudoscalar mesons become  decoupled from the properties of glueballs (hence, independent of condensate $h_0$). However, once the interactions of quark components with glueballs are turned on the predictions are expected to depend on $h_0$.  The predictions for both scalar and pseudoscalar SU(3) singlets include a light and a heavy state.   The light  pseudoscalar singlet mass of 93 MeV does not overlap with known physical eta masses, but the scalar singlet mass of 236 MeV is  qualitatively pointing to the sigma meson.   The lighter than expected pseudoscalar singlet can be attributed to the complete suppression of pseudoscalar glueball in this limit which in turn suppresses the realization of U(1)$_{\rm A}$ which is known to be crucial in generating the correct $\eta$ masses.  
The low mass of the $\eta$ meson in our method can be justified further. In \cite{Weinberg75} Weinberg showed that in a theory where  U(1)$_{\rm A}$ axial current is conserved and its breaking stems only from the quark masses there is a pseudoscalar singlet with a very low mass of order $\sqrt{3}\, m_{\pi}$.  Similarly,  in \cite{ChUA7} within the leading-order U(3) chiral perturbation theory,  an explicit calculation showed that in the $N_c \rightarrow \infty$ limit (where the axial anomaly term vanishes) the eta mass approaches the mass of the pion.

For the scalar singlet the situation is different because the effect of trace anomaly on singlet scalar masses is not as pronounced as the effect of axial anomaly on singlet pseudoscalars \cite{Jora1}-\cite{Jora7}.    The heavier singlet masses (both around 1.5 GeV) overlap with some of the known $\eta$ or $f_0$ states above 1 GeV.

The prediction for the scalar glueball mass depends on $h_0$. We have shown \cite{Jora25} that the favored  range of  $h_0 = 0.8-1$ is consistent with QCD sum-rules analysis.  In this range of $h_0$ the scalar glueball mass is 1.6-2.0 GeV  consistent with other approaches \cite{Huang}.

In the decoupling limit,  the model predictions for the quark and glue contents of scalars and pseudoscalars  are expected to be of qualitative importance. We have presented these predictions in Table \ref{T_Rotation_matrices}.   We see that the substructure of octet states (both scalars and pseudoscalars) are overall consistent with general expectations where light pseudoscalars are mainly  quark-antiquark states (which is seen to be minimally the case) while light scalars are closer to four-quark states (which is seen to be clearly the case).    The situation for singlets is different and the predictions are not conclusive because of suppressing the interactions of quark components with glueballs.   The values of $\alpha$, $\beta$ and $A$ are independent of the parameter $h_0$.

\begin{table}[htbp]
	\begin{center}
		\caption{Values of the model parameters in terms of $h_0$.}
		\begin{tabular}{c|c|c}
			{\rm Parameters} &$h_0=0.80$\,\,${\rm GeV}$&$h_0=1.0$\,\,${\rm GeV}$\\
			\hline
			\hline
			$\alpha$    (${\rm GeV}$)		&	$5.10 \times 10 ^{-2}$&   $5.10 \times 10 ^{-2}$\\
			$\beta$ (${\rm GeV}$)           &$4.12 \times 10^{-2}$ &  $4.12 \times 10^{-2}$  \\
			$u_1$&$1.83 \times 10 $&$2.85\times 10 $\\
			$u_2$&$-1.06 \times 10^{3}$&$-2.68 \times 10^{3}$\\
			$u_3$&$8.03 \times 10^{-1} $&$5.14 \times 10 ^{-1}$\\
			$u_4$&$-3.39$&$-2.72$\\
			$\lambda_2$&$-3.56 \times 10^{-2}$&$-3.65 \times 10^{-2}$\\
			$\lambda_3$&$-2.13\times 10^{-3}$&$-8.71\times 10^{-4}$\\
			$A$\,\, $({\rm GeV^3})$&$7.96\times 10^{-4}$&$7.96\times 10^{-4}$\\
			\hline
			\hline	
		\end{tabular}
		\label{T_param}
	\end{center}
\end{table}

\begin{table}[htbp]
	\begin{center}
		\caption{Values of the model parameters in terms of $h_0$.}
		\begin{tabular}{c|c|c}
			{\rm Masses (GeV)} &$h_0=0.80$\,\,${\rm GeV}$&$h_0=1.0$\,\,${\rm GeV}$\\
			\hline
			\hline
$m_{0^-}$ 	     & 9.26 	$\times 10^{-2}$   &   9.26 	$\times 10^{-2}$ \\
${m'}_{0^-}$     & 1.58                        &  1.58 \\
$m_{0^+}$        & $2.36  \times 10^{-1}$     &  $2.36  \times 10^{-1}$\\
${m'}_{0^+}$     &  1.52                &1.52 \\
$m_h$            &  1.60                &2.0\\
			\hline
			\hline	
		\end{tabular}
			\label{T_SU3_singlet_masses}
	\end{center}
\end{table}

\begin{table}[htbp]
	\begin{center}
		\caption{Values of the model parameters in terms of $h_0$.}
		\begin{tabular}{c|c|c}
					{\rm Masses (GeV)} &$h_0=0.80$\,\,${\rm GeV}$&$h_0=1.0$\,\,${\rm GeV}$\\
			\hline
			\hline
$\left[{K_{8^-}}\right]^{-1}$     &	
$\begin{array}{cc}
0.772   & 0.636 \\
-0.636  & 0.772
\end{array}
$
&
$\begin{array}{cc}
0.772  &   0.636 \\
-0.636 &   0.772
\end{array}
$
\\
\hline
$\left[{{\widehat K}_{0^-}}\right]^{-1}$     &	
$\begin{array}{cc}
0.521  &  -0.853 \\
-0.853 &  -0.521
\end{array}
$
&
$
\begin{array}{cc}
0.521  &  -0.853 \\
-0.853 &  -0.521
\end{array}
$
\\
\hline
$\left[{K_{8^+}}\right]^{-1}$     &	
$\begin{array}{cc}
0.235  &  -0.972 \\
-0.972 &  -0.235
\end{array}
$
&
$
\begin{array}{cc}
0.235   &  -0.972 \\
-0.972  &  -0.235
\end{array}
$
\\
\hline
$\left[{{\widehat K}_{0^+}}\right]^{-1}$     &	
$\begin{array}{cc}
0.765  &  0.644 \\
-0.644 &  0.765
\end{array}
$
&
$
\begin{array}{cc}
0.765  &   0.644 \\
-0.644 &  0.765
\end{array}
$
\\
\hline
\hline	
		\end{tabular}
	\label{T_Rotation_matrices}
	\end{center}
\end{table}

To further investigate the stability of predictions, we have considered the decoupling limit with massless quark.    This imposes more stringent conditions on the system of equations.   The results are given in the three tables of Appendix B and show, expectedly,  that there are no sensitivities when this additional condition is imposed.

According to Eq. (\ref{etama5466}) in order to decouple the pseudoscalar glueball (which amounts to setting the elements $(M^2_{\eta})_{i5}$ with $i=1,2,3,4$ to zero) the instanton term should be dropped altogether. However alternatively one might consider a generalized instanton term of the type in Eq. (\ref{potential73568}) that contains enough parameters such that the decoupling equations are solvable and the axial anomaly would still be satisfied.

We end this section by a general comparison of the decoupling limit presented here and the decoupling of scalar glueballs from quark-antiquarks obtained as a result of a ``chiral suppression'' studied in \cite{Chanowitz:2005du}.  Of course there are major differences between our formulation and the work of \cite{Chanowitz:2005du}:   Our framework is formulated in terms of chiral nonet fields and can only indirectly probe quarks (through studies of mixing patterns among components);  it contains composites of four-quark fields; and in our model chiral symmetry is broken both through quark masses as well as spontaneously through condensates of quark-antiquarks, four-quarks and scalar glueball field.  Moreover, the decoupling limit in our model can occur both with or without quark masses,  and in either case,  within the SU(3)$_{\rm V}$ subgroup.  In this decoupling limit,  while the scalar glueball is not interacting with mesons, it still plays a very important role in stabilizing the vacuum through mixing  of its condensate with other condensates.   Nevertheless,  in limits that our model  qualitatively resembles  the model of Ref. \cite{Chanowitz:2005du}, it does not seem to contradict the ``chiral suppression'' found in that work.  For example,  when trace anomaly is saturated by glueballs only (i.e. when 
$\lambda_2=\lambda_3=0$) and assuming $\gamma_m=2$ for simplicity, together with invoking the minimum conditions (\ref{res5678}),  the coupling of scalar glueball in this limit to non-strange quark-antiquark is:
\begin{equation}
g_{n{\bar n}} \approx {1\over \sqrt{2}}\,
\lim_{\lambda_2=\lambda_3\rightarrow 0, 
\gamma_m \rightarrow 2} \left(X_0^2\right)_{15}= -4  \,
\left(
{\frac 
{{ A_1}\,{\alpha_3}^{3}-{ A_3}\,{\alpha_1}^{3}  }{{ h_0}\, 
\left( {\alpha_1}^{2}-{\alpha_3}^{2}
 \right) \alpha_3}}
\right)
 \label{E_chiral_sup_nn}
\end{equation}
whereas  the coupling of scalar glueball to strange quark-antiquark (in this limit) becomes:
\begin{equation}
g_{s{\bar s}} \approx
 \lim_{\lambda_2=\lambda_3\rightarrow 0, 
\gamma_m \rightarrow 2} \left(X_0^2\right)_{25}= -4  \, 
\left(
{\frac 
{{ A_1}\,{\alpha_3}^{3}-{ A_3}\,{\alpha_1}^{3}  }{{ h_0}\, 
\left( {\alpha_1}^{2}-{\alpha_3}^{2}
 \right) \alpha_1}}
\right)
 \label{E_chiral_sup_ss}
\end{equation}
First, we see that in massless quark limit, both of these couplings vanish in agreement with ``chiral suppression'' of 
\cite{Chanowitz:2005du}.   Secondly, using (\ref{E_chiral_sup_nn}) and (\ref{E_chiral_sup_ss}) we find:
\begin{equation}
{g_{ss}\over g_{nn}} \approx {\alpha_3\over \alpha_1}.
\end{equation}
To get the numerical value of this ratio, we need to do the calculation in the SU(2) isospin limit.   In the absence of the SU(2) calculation in the present work, we can only give a rough approximation.  For all values of the condensates $\alpha_1$ and $\alpha_3$ found in the generalized linear sigma model without glueballs (see Fig. 2 in Ref. \cite{Jora5})  this ratio is clearly larger than one in agreement with the ``chiral suppression'' of  \cite{Chanowitz:2005du}.    However, this is only an estimate and has to be confirmed with full SU(2) calculation.

\section{Conclusions}

In \cite{Jora41,Jora42,Jora43} we introduced the SU(3)$_{\rm V}$ limit of the generalized linear sigma model with two chiral nonets (both with and without quark masses). These analyses were performed with the hope that this simplified model would have the correct main features of the low energy meson spectrum. In the limit of massless quarks three pions and one of the $\eta$'s were massless as expected. The model contained a very low isosinglet scalar mass and was consistent with the picture developed in further works that the low lying pseudoscalars are mostly ``quark-antiquark" states whereas the low lying scalars have a larger four quark component. The SU(3)$_{\rm V}$ limit in the presence of quark masses had the same main features except that the massless states were replaced by light meson masses.

In the model presented here all the characteristics outlined in our previous versions of the generalized linear model are respected with the provision that  one of the pseudoscalar singlets and the lowest scalar isosinglet ($\sigma$) are nearly  massless in the limit of massless quarks. This difference stems from the presence of the  instanton term in the  Lagrangian in the model discussed in \cite{Jora41,Jora42,Jora43} that brought a large contribution directly to the $\eta$ masses and indirectly to the $\sigma$ mass. Since the corresponding term gives no contribution to the $\eta$ mass matrix in the decoupling limit of the present model,  it is natural to obtain a massless $\eta$. This set-up may have limited phenomenological consequences but it is very important from the theoretical point of view as it reveals  the significance of specific terms in the Lagrangian. Also the important feature that the low lying pseudoscalars are mainly ``quark-antiquark" states whereas the corresponding scalars have large four quark components emerges from the rotation matrices in Tables \ref{T_Rotation_matrices} and \ref{T_Rotation_matrices_massless}.

In the case of massive quarks in the decoupling SU(3)$_{\rm V}$ limit there is no massless meson but  the main characteristics of the meson spectrum are preserved. The model contains a very low $\eta$ and a very low $\sigma$ with the masses indicated in Table \ref{T_SU3_singlet_masses}.

The generalized linear sigma model developed in \cite{Jora1}-\cite{Jora5} provided an adequate and reliable picture of low-energy QCD model with pseudoscalar and scalar mesons. In the present work we expanded that framework  to accommodate scalar and pseudoscalar glueballs that may mix with the quark meson states and conceivably  lead to a better description of the low energy sector. We discussed that in the decoupling SU(3)$_{\rm V}$ limit (both with and without  quark masses)  the model presents interesting features compatible with the same limit for the generalized linear sigma model of quark mesons only. Moreover,  the mass of the scalar glueball was predicted for adequate values of the glueball condensate with a result in very good agreement with those calculated in lattice studies \cite{Bali1,Morningstar,Chen} or from QCD sum rules \cite{Huang} (for a detailed discussion of the scalar glueball mass in the present approach see \cite{Jora25}).

The Lagrangian in (\ref{inlgr567}) is fairly complex, contains a very large number of parameters that, in principle,  can be exactly treated numerically.
However, in practice, brute force numerical approach is insufficient because (i) it lacks physical insight into how the system of equations evolve step by step from the simplest limits which are exactly solvable and contain the basic  fundamental knowledge of the underlying dynamics, and (ii)
the optimized numerical solutions of a highly nonlinear system of equations with many unknown parameters (in the leading order of the model presented here there are 18 a priori unknown parameters) are not unique and while may mathematically correspond to an optimized solution may not necessarily correspond to a physical solution.   Therefore,  tackling the system of equations by first pushing them to solvable limits and then step by step evolving their solutions to the desired general conditions is necessary and is the main strategy promoted in this work for exploring the unknowns of the model.
 The aspects already discussed here are promising and  suggest that a comprehensive analysis of the model may lead to a very interesting picture of low energy scalar and pseudoscalar meson spectrum and properties. This  endeavor  will be further pursued in future works.

\appendix

\section{Isosinglet mass matrices}
In this appendix we give the mass matrices for isosinglet scalars and isosinglet pseudoscalars:

\begin{eqnarray}
\left(X^2_0\right)_{11} &=&
4\,{\it u_4}\,{\it h_0}\,\beta_3+2\,{\it u_1}\,{{\it h_0}}^{2}+12\,{\it u_2}
\,{\alpha_1}^{2}
-8\,{\frac
	{
		{\it h_0}^{4}
		\left(
		\left(\beta_1\,\alpha_1 + {1\over 2} \alpha_3\,\beta_3 \right)^2 \,\lambda_2
		+ {1\over 2} \, \alpha_1^2\, \beta_1^2 \lambda_3
		\right)
	}
	{
		\alpha_1^{2} \left( 2\,\beta_1\,\alpha_1+\beta_3\,\alpha_3 \right) ^{2}
	}
}
\nonumber\\
&&
-
4\, \left( 2\,{\alpha_1}^{2}+{\alpha_3}^{2} \right) ^{-1-{\it \gamma_m
	}/2} \left( -2+{\it \gamma_m} \right) ^{2}{\alpha_1}^{2} \left( 2\,{
	\it A_1}\,\alpha_1+{\it A_3}\,\alpha_3 \right)
\nonumber\\
&&
+4\, \left( -2+{\it
	\gamma_m} \right)  \left( {\it A_1}\,\alpha_1+1/2\,{\it A_3}\,\alpha_3
\right)  \left( 2\,{\alpha_1}^{2}+{\alpha_3}^{2} \right) ^{-{\it
		\gamma_m}/2}
\nonumber\\
&&
-16\, \left( -2+{\it \gamma_m} \right)  \left( 2\,{
	\alpha_1}^{2}+{\alpha_3}^{2} \right) ^{-1-{\it \gamma_m}/2}{\alpha_1}^{
	2} \left( {\it A_1}\,\alpha_1+1/2\,{\it A_3}\,\alpha_3 \right)
\nonumber\\
&&
+8\,
\left( 2\,{\alpha_1}^{2}+{\alpha_3}^{2} \right) ^{-{\it \gamma_m}/2}
\left( -2+{\it \gamma_m} \right) \alpha_1\,{\it A_1}
\nonumber\\
\left(X^2_0\right)_{12} &=&
-
{1\over
	{ \left( 2\,\beta_1\,\alpha_1+\beta_3\,\alpha_3
		\right) ^{2} \left( 2\,{\alpha_1}^{2}+{\alpha_3}^{2} \right) ^{2}}
}
\nonumber\\
&&
\times\left[
16\,
\left(  \left( -2+{\it \gamma_m} \right)  \left( -{
	\it A_3}\,{\alpha_1}^{3}+{\it A_1}\,\alpha_3\, \left( {\it \gamma_m}-1
\right) {\alpha_1}^{2}
+1/2\,{\it A_3}\,{\alpha_3}^{2} \left( {\it
	\gamma_m}-1 \right) \alpha_1-1/2\,{\it A_1}\,{\alpha_3}^{3} \right)
\right.
\right.
\nonumber\\
&&		
\left( \beta_1\,\alpha_1+1/2\,\beta_3\,\alpha_3 \right) ^{2} \left( 2\,{
	\alpha_1}^{2}+{\alpha_3}^{2} \right) ^{1-{\it \gamma_m}/2}-1/2\,
\left( {\alpha_1}^{2}+1/2\,{\alpha_3}^{2} \right) ^{2}
\nonumber\\
&&
\left.
\left.
\left( 8\,{\it
	u_4}\,{\beta_1}^{2}{\alpha_1}^{2}+8\,{\it u_4}\,\beta_1\,\beta_3\,\alpha_1\,
\alpha_3+ \left( 2\,{\it u_4}\,\beta_3\,{\alpha_3}^{2} -\lambda_3 {{\it h_0}}^{3} \right) \beta_3 \right) {\it h_0}\,
\beta_1 \right) \sqrt {2}
\right]
\nonumber\\
\left(X^2_0\right)_{13} &=&
4\,{\it u_4}\,{\it h_0}\,\alpha_3+{{\it h_0}}^{4} \left( 2\,{\frac {\lambda_3}{2\,\beta_1\,\alpha_1+\beta_3\,\alpha_3}}-4\,{\frac {
		\lambda_3  \beta_1\,\alpha_1}{ \left( 2\,\beta_1
		\,\alpha_1+\beta_3\,\alpha_3 \right) ^{2}}} \right)
\nonumber\\
\left(X^2_0\right)_{14} &=&
{\it u_4}\,{\it h_0}\, \left( -4\,\sqrt {2}\alpha_1-4\,\alpha_3\,\sqrt {2}
+4\, \left( 2\,\alpha_1+\alpha_3 \right) \sqrt {2} \right) -2\,{\frac {{
			{\it h_0}}^{4}  \lambda_3 \sqrt {2}\beta_1\,
		\alpha_3}{ \left( 2\,\beta_1\,\alpha_1+\beta_3\,\alpha_3 \right) ^{2}}}
\nonumber\\
\left(X^2_0\right)_{15} &=&
{\it u_4}\, \left( 8\,\sqrt {2}\beta_1\,\alpha_1-4\, \left( 2\,\beta_1+
\beta_3 \right) \sqrt {2}\alpha_1-4\,\sqrt {2}\beta_1\, \left( 2\,\alpha_1
+\alpha_3 \right) -4\, \left( 2\,\beta_1\,\alpha_1+\beta_3\,\alpha_3
\right) \sqrt {2}
+4\, \left( 2\,\beta_1+\beta_3 \right)  \left( 2\,
\alpha_1+\alpha_3 \right) \sqrt {2} \right)
\nonumber\\
&&
+4\,{\it u_1}\,{\it h_0}\,
\sqrt {2}\alpha_1+4\,{{\it h_0}}^{3} \left( 2\,{\frac {\lambda_2\,\sqrt {
			2}}{\alpha_1}}+2\,{\frac {  \lambda_3  \sqrt {2}
		\beta_1}{2\,\beta_1\,\alpha_1+\beta_3\,\alpha_3}} \right)
\nonumber\\
\left(X^2_0\right)_{22} &=&
{1\over 	{{\alpha_3}^{2} \left( 2\,\beta_1\,\alpha_1+\beta_3\,\alpha_3
		\right) ^{2}}}
\left( -16\,{\alpha_3}^{4} \left( {\it A_1}\,\alpha_1+1/2\,{\it A_3}\,
\alpha_3 \right)  \left( \beta_1\,\alpha_1+1/2\,\beta_3\,\alpha_3 \right) ^
{2}{\it \gamma_m}\, \left( -2+{\it \gamma_m} \right)
\right.
\nonumber\\
&&		
\left( 2\,{
	\alpha_1}^{2}+{\alpha_3}^{2} \right) ^{-1-{\it \gamma_m}/2}+16\,{
	\alpha_3}^{2} \left( \beta_1\,\alpha_1+1/2\,\beta_3\,\alpha_3 \right) ^{2}
\left( {\it A_1}\,\alpha_1+3/2\,{\it A_3}\,\alpha_3 \right)  \left( -2+{
	\it \gamma_m} \right)  \left( 2\,{\alpha_1}^{2}+{\alpha_3}^{2}
\right) ^{-{\it \gamma_m}/2}
\nonumber\\
&&		
+12\,{\alpha_3}^{6}{\beta_3}^{2}{\it u_2}+
48\,\alpha_1\,{\alpha_3}^{5}\beta_1\,\beta_3\,{\it u_2}+ \left( 48\,{\alpha_1}^{2}{\beta_1}^{2}{\it u_2}+2\,{\beta_3}^{2}{{\it h_0}}^{2}{\it u_1}
\right) {\alpha_3}^{4}+8\,\alpha_1\,{\alpha_3}^{3}\beta_1\,\beta_3\,{{\it
		h_0}}^{2}{\it u_1}
\nonumber\\
&&		
\left.				
-2\,{{\it h_0}}^{2} \left( -4\,{\it u_1}\,{\beta_1}^{2}{
	\alpha_1}^{2}+{{\it h_0}}^{2}{\beta_3}^{2} \left( \lambda_2+\lambda_3
\right)  \right) {\alpha_3}^{2}-8\,\alpha_1\,\alpha_3\,\beta_1\,\beta_3\,{
	{\it h_0}}^{4}\lambda_2-8\,{\alpha_1}^{2}{\beta_1}^{2}{{\it h_0}}^{4}
\lambda_2
\right)
\nonumber\\
\left(X^2_0\right)_{23} &=&
{\it u_4}\,{\it h_0}\, \left( -4\,\sqrt {2}\alpha_1-4\,\alpha_3\,\sqrt {2}
+4\, \left( 2\,\alpha_1+\alpha_3 \right) \sqrt {2} \right) -2\,{\frac {{
			{\it h_0}}^{4}  \lambda_3 \beta_3\,\sqrt {2}
		\alpha_1}{ \left( 2\,\beta_1\,\alpha_1+\beta_3\,\alpha_3 \right) ^{2}}}
\nonumber\\
\left(X^2_0\right)_{24} &=&
{{\it h_0}}^{4} \left( 2\,{\frac {\lambda_3}{2\,\beta_1\,
		\alpha_1+\beta_3\,\alpha_3}}-2\,{\frac { \lambda_3 \beta_3\,\alpha_3}{ \left( 2\,\beta_1\,\alpha_1+\beta_3\,\alpha_3
		\right) ^{2}}} \right)
\nonumber\\
\left(X^2_0\right)_{25} &=&
{\it u_4}\, \left( 4\,\beta_3\,\alpha_3-4\, \left( 2\,\beta_1+\beta_3
\right) \alpha_3-4\,\beta_3\, \left( 2\,\alpha_1+\alpha_3 \right) -8\,
\beta_1\,\alpha_1+4\, \left( 2\,\beta_1+\beta_3 \right)  \left( 2\,\alpha_1
+\alpha_3 \right)  \right) +4\,{\it u_1}\,{\it h_0}\,\alpha_3
\nonumber\\
&&
+4\,{{\it h_0}
}^{3} \left( 2\,{\frac {\lambda_2}{\alpha_3}}+2\,{\frac { \lambda_3 \beta_3}{2\,\beta_1\,\alpha_1+\beta_3\,\alpha_3}}
\right)
\nonumber\\
\left(X^2_0\right)_{33} &=&
2\,{\it u_3}\,{{\it h_0}}^{2}-4\,{\frac {{{\it h_0}}^{4} \lambda_3 {\alpha_1}^{2}}{ \left( 2\,\beta_1\,\alpha_1+\beta_3\,
		\alpha_3 \right) ^{2}}}
\nonumber\\
\left(X^2_0\right)_{34} &=&
-2\,{\frac {{{\it h_0}}^{4} \lambda_3 \sqrt {2
		}\alpha_1\,\alpha_3}{ \left( 2\,\beta_1\,\alpha_1+\beta_3\,\alpha_3 \right)
		^{2}}}
\nonumber\\
\left(X^2_0\right)_{35} &=&
{\it u_4}\, \left( 4\,\sqrt {2}{\alpha_1}^{2}-2\,\sqrt {2} \left( 2\,{
	\alpha_1}^{2}+{\alpha_3}^{2} \right) -4\,\sqrt {2}\alpha_1\, \left( 2\,
\alpha_1+\alpha_3 \right) +2\,\sqrt {2} \left( 2\,\alpha_1+\alpha_3
\right) ^{2} \right) +4\,{\it u_3}\,{\it h_0}\,\sqrt {2}\beta_1
\nonumber\\
&&
+8\,{
	\frac {{{\it h_0}}^{3} \lambda_3 \sqrt {2}
		\alpha_1}{2\,\beta_1\,\alpha_1+\beta_3\,\alpha_3}}
\nonumber\\
\left(X^2_0\right)_{44} &=&
2\,{\it u_3}\,{{\it h_0}}^{2}-2\,{\frac {{{\it h_0}}^{4} \lambda_3 {\alpha_3}^{2}}{ \left( 2\,\beta_1\,\alpha_1+\beta_3\,
		\alpha_3 \right) ^{2}}}
\nonumber\\
\left(X^2_0\right)_{45} &=&
{\it u_4}\, \left( 2\,{\alpha_3}^{2}-4\,{\alpha_1}^{2}-4\,\alpha_3\,
\left( 2\,\alpha_1+\alpha_3 \right) +2\, \left( 2\,\alpha_1+\alpha_3
\right) ^{2} \right) +4\,{\it u_3}\,{\it h_0}\,\beta_3+8\,{\frac {{{\it
				h_0}}^{3} \lambda_3 \alpha_3}{2\,\beta_1\,\alpha_1+\beta_3\,\alpha_3}}
\nonumber\\
\left(X^2_0\right)_{55} &=&
2\,{\it u_1}\, \left( 2\,{\alpha_1}^{2}+{\alpha_3}^{2} \right) +2\,{\it
	u_3}\, \left( 2\,{\beta_1}^{2}+{\beta_3}^{2} \right) +12\,{\it u_5}\,{{
		\it h_0}}^{2}
\nonumber\\
&&
+12\,{{\it h_0}}^{2} \left( \lambda_1\,\ln  \left( {\frac {{
			{\it h_0}}^{4}}{{\Lambda}^{4}}} \right) +2\,\lambda_2\,\ln  \left( {
	\frac {{\alpha_1}^{2}\alpha_3}{{\Lambda}^{3}}} \right) +2\, \lambda_3 \ln  \left( {\frac {2\,\beta_1\,\alpha_1+\beta_3\,\alpha_3}{{\Lambda}^{2}}} \right)  \right) +28\,{{\it h_0}}^{2}
\lambda_1
\end{eqnarray}

\begin{eqnarray}
\left(M^2_\eta\right)_{11} &=&
-4\,{\it u_4}\,{\it h_0}\,\beta_3+2\,{\it u_1}\,{{\it h_0}}^{2}+4\,{\it u_2}
\,{\alpha_1}^{2}
+ 8\,{\frac { \left(  \left( \beta_1\,\alpha_1+1/2\,\beta_3\,\alpha_3
		\right) ^{2}\lambda_2+1/2\,\lambda_3\,{\beta_1}^{2}{\alpha_1}^{2}
		\right) {{\it h_0}}^{4}}{{\alpha_1}^{2} \left( 2\,\beta_1\,\alpha_1+\beta_3\,\alpha_3 \right) ^{2}}}
\nonumber\\
&&
+
4\, \left( 2\,{\alpha_1}^{2}+{\alpha_3}^{2} \right) ^{-{\it \gamma_m}/
	2} \left( -2+{\it \gamma_m} \right)  \left( {\it A_1}\,\alpha_1+1/2\,{
	\it A_3}\,\alpha3 \right)
\nonumber\\
\left(M^2_\eta\right)_{12} &=&
{\it u_4}\,{\it h_0}\, \left( 4\,\sqrt {2}\beta_1+4\,\beta_3\,\sqrt {2}-4
\, \left( 2\,\beta_1+\beta_3 \right) \sqrt {2} \right) +2\,{\frac {{{
				\it h_0}}^{4}\lambda_3\,\sqrt {2}\beta_1\,\beta_3}{ \left( 2\,\beta_1\,
		\alpha_1+\beta_3\,\alpha_3 \right) ^{2}}}
\nonumber\\
\left(M^2_\eta\right)_{13} &=&
-4\,{\it u_4}\,{\it h_0}\,\alpha_3+{{\it h_0}}^{4}\lambda_3\, \left(
{2\over
	{ 2\,\beta_1\,\alpha_1+\beta_3\,\alpha_3 }
}
-4\,{\frac {
		\beta_1\,\alpha_1}{ \left( 2\,\beta_1\,\alpha_1+\beta_3\,\alpha_3 \right) ^{
			2}}} \right)
\nonumber\\
\left(M^2_\eta\right)_{14} &=&
{\it u_4}\,{\it h_0}\, \left( 4\,\sqrt {2}\alpha_1+4\,\alpha_3\,\sqrt {2}-
4\, \left( 2\,\alpha_1+\alpha_3 \right) \sqrt {2} \right) -2\,{\frac {{{
				\it h_0}}^{4}\lambda_3\,\sqrt {2}\beta_1\,\alpha_3}{ \left( 2\,\beta_1\,
		\alpha_1+\beta_3\,\alpha_3 \right) ^{2}}}
\nonumber\\
\left(M^2_\eta\right)_{15} &=&
{\frac { \left( \beta_1\, \left( \gamma_1+1 \right) \alpha_1+\alpha_3\,
		\beta_3\,\gamma_1 \right) \sqrt {2}{{\it h_0}}^{3}}{12\,{\alpha_1}^{2}
		\beta_1+6\,\alpha_1\,\alpha_3\,\beta_3}}
\nonumber\\
\left(M^2_\eta\right)_{22} &=&
2\,{\it u_1}\,{{\it h_0}}^{2}+4\,{\it u_2}\,{\alpha_3}^{2}
+8\,{\frac {{{\it h_0}}^{4} \left(  \left( \beta_1\,\alpha_1+1/2\,\beta_3\,
		\alpha_3 \right) ^{2}\lambda_2+1/4\,\lambda_3\,{\beta_3}^{2}{\alpha_3}^{2}
		\right) }{{\alpha_3}^{2} \left( 2\,\beta_1\,\alpha_1+\beta_3\,\alpha_3
		\right) ^{2}}}
\nonumber\\
&&
+4\, \left( 2\,{\alpha_1}^{2}+{\alpha_3}^
{2} \right) ^{-{\it \gamma_m}/2} \left( -2+{\it \gamma_m} \right)
\left( {\it A_1}\,\alpha_1+1/2\,{\it A_3}\,\alpha_3 \right)
\nonumber\\
\left(M^2_\eta\right)_{23} &=&
-2\,{\frac {{\it h_0}\,\sqrt {2}\alpha_1\, \left( 8\,{\it u_4}\,{\beta_1}^
		{2}{\alpha_1}^{2}+8\,{\it u_4}\,\beta_1\,\beta_3\,\alpha_3\,\alpha_1+2\,{
			\alpha_3}^{2}{\beta_3}^{2}{\it u_4}+\beta_3\,{{\it h_0}}^{3}\lambda_3
		\right) }{ \left( 2\,\beta_1\,\alpha_1+\beta_3\,\alpha_3 \right) ^{2}}}
\nonumber\\
\left(M^2_\eta\right)_{24} &=&
4\,{\frac {\alpha_1\,\beta_1\,{{\it h_0}}^{4}\lambda_3}{ \left( 2\,\beta_1
		\,\alpha_1+\beta_3\,\alpha_3 \right) ^{2}}}
\nonumber\\
\left(M^2_\eta\right)_{25} &=&
{\frac { \left( 2\beta_1\,  \alpha_1\gamma_1+\alpha_3\,
		\beta_3\, \right) {{\it h_0}}^{3}}{12\,\alpha_1\alpha_3\beta_1
		+6\,\alpha_3^2\,\beta_3}}
\nonumber\\
\left(M^2_\eta\right)_{33} &=&
2\,{\frac {{{\it h_0}}^{2} \left( 4\,{\alpha_1}^{2}{\beta_1}^{2}{\it u_3}+
		2\,{\alpha_1}^{2}{{\it h_0}}^{2}\lambda_3+4\,\alpha_1\,\alpha_3\,\beta_1\,
		\beta_3\,{\it u_3}+{\alpha_3}^{2}{\beta_3}^{2}{\it u_3} \right) }{ \left( 2
		\,\beta_1\,\alpha_1+\beta_3\,\alpha_3 \right) ^{2}}}
\nonumber\\
\left(M^2_\eta\right)_{34} &=&
2\,{\frac { \lambda_3 {{\it h_0}}^{4}\sqrt {
			2}\alpha_1\,\alpha_3}{ \left( 2\,\beta_1\,\alpha_1+\beta_3\,\alpha_3
		\right) ^{2}}}
\nonumber\\
\left(M^2_\eta\right)_{35} &=&
{\frac {  \alpha_1\, \left( \gamma_1-1 \right) \sqrt {2}{{\it h_0}}^{3}}{12\,\alpha_1
		\beta_1+6\,\alpha_3\,\beta_3}}
\nonumber\\
\left(M^2_\eta\right)_{44} &=&
2\,{\frac {{{\it h_0}}^{2} \left( 4\,{\alpha_1}^{2}{\beta_1}^{2}{\it u_3}+
		4\,\alpha_1\,\alpha_3\,\beta_1\,\beta_3\,{\it u_3}+{\alpha_3}^{2}{\beta_3}^{2
		}{\it u_3}+{\alpha_3}^{2}{{\it h_0}}^{2}\lambda_3 \right) }{ \left( 2\,
		\beta_1\,\alpha_1+\beta_3\,\alpha_3 \right) ^{2}}}
\nonumber\\
\left(M^2_\eta\right)_{45} &=&
{\frac {  \alpha_3\, \left( \gamma_1-1 \right) {{\it h_0}}^{3}}{12\,\alpha_1
		\beta_1+6\,\alpha_3\,\beta_3}}
\nonumber\\
\left(M^2_\eta\right)_{55} &=&
2\,{\it u_6}\,{{\it h_0}}^{2}
\label{etama5466}
\end{eqnarray}
Using the mass matrix for the eta system, together with vacuum conditions  (\ref{res5678}), we can show 
\begin{eqnarray}
\det\left(M^2_\eta\right)&=&
1/6\,
{\frac { \left( {\alpha_1}^{2}+1/2\,{\alpha_3}^{2} \right)  \left( 
\alpha_1+\alpha_3 \right) \lambda_3\,{{\it h_0}}^{10} \left( -\alpha_3+
\alpha_1 \right) }{ \left( \beta_1\,\alpha_1-\beta_3\,\alpha_3 \right) ^{2}
\alpha_3\, \left( \beta_1\,\alpha_1+1/2\,\beta_3\,\alpha_3 \right) ^{3}
\alpha_1}} 
\nonumber \\
&&
\Bigg[ -144\,{A_1}\, A_3
          \left( 2\,{\alpha_1}^{2}+\, {\alpha_3}^{2}\right) ^{1 -{\it \gamma_m}}
          \left( 2\,{\alpha_1}^{2}+\, {\alpha_3}^{2}\right)  
          \left( -{\frac {\beta_1\, \left( -1+\gamma_1 \right)^{2}
                {\alpha_1}^{3}}{216}}-
\right.
\nonumber \\
&&               
\alpha_3\, 
          \left( \lambda_3\,{\it u_6}-{\frac {{\gamma_1}^{2}}{216}}+{\frac {\gamma_1}{108}}-{\frac{1}{216}} \right) 
\beta_3\,{\alpha_1}^{2}
\nonumber\\
&&
\left.
+\beta_1\,{\alpha_3}^{2} 
\left( 
       \lambda_3\,{\it u_6}-{\frac {{\gamma_1}^{2}}{432}}+{\frac {\gamma_1}{216}}-{\frac{1}{432}}
\right) \alpha_1
        +{\frac {\beta_3\,{\alpha_3}^{3} \left( -1+\gamma_1\right) ^{2}}{432}} 
\right)  
 \nonumber\\
&&
 +\lambda_3\,{{h_0}}^{4} \left( 
-\alpha_1\,\beta_3+\beta_1\,\alpha_3 \right)  \left( {A_1}\,\alpha_1+2\,
{A_3}\,\alpha_3 \right)  \left( 2\,{\alpha_1}^{2}+{\alpha_3}^{2}
 \right) ^{-{\gamma_m}/2} 
\Bigg], 
\end{eqnarray}
which, as expected, shows: 
\begin{equation}
\lim_{A_1,A_3\rightarrow 0 }\det\left(M^2_\eta\right) =  0.
\end{equation}

\section{Decoupling limit in the massless case}
\label{A_massless}

In this appendix we consider the decoupling limit in the absence of explicit symmetry breaking term in the Lagrangian.
This amounts to the condition $A_1=A_3=A=0$ where equivalently from the first and second equations in (\ref{res5678}):
\begin{eqnarray}
6u_4h_0\alpha\beta+4u_2\alpha^3+u_1h_0^2\alpha=0.
\label{massless657cond}
\end{eqnarray}
Equations (\ref{lambda3_sol}), (\ref{res674665}), (\ref{u3_sol}) and (\ref{lambda2_sol}) are still valid.
The system of equations,
\begin{eqnarray}
\frac{1}{4}(1-6\lambda_2-4\lambda_3)&=& \lambda_1 = \frac{11}{36}
\nonumber\\
{\rm Tr}
 \left[N_8^2\right] &=& \left(m'_{8^-}\right)^2
\nonumber\\
\det \left[N_8^2\right]&=& 0,
\label{ystnew53888}
\end{eqnarray}
is solved for the parameters $u_1$, $u_2$, $u_4$:
\begin{eqnarray}
&&u_1=\frac{2h_0^2}{27\alpha^2},
\nonumber\\
&&u_2=-\frac{2\alpha^2h_0^4+2\beta^2h_0^4-27\alpha^2\beta^2  \left({m'}_{8^-}\right)^2  }{108\alpha^4(\alpha^2+\beta^2)},
\nonumber \\
&&u_4=-\frac{\beta  \left({m'}_{8^-}\right)^2 }{6(\alpha^2+\beta^2)h_0}.
\label{someparamry65}
\end{eqnarray}
Furthermore, we calculate $\alpha$ from  (see \cite{Jora41,Jora42}),
\begin{eqnarray}
\alpha=\frac{f_{\pi}\sqrt{3  \left({m'}_{8^-}\right)^2  - \left({m}_{8^+}\right)^2 - \left({m'}_{8^+}\right)^2}}
{2\sqrt{2}\left({m'}_{8^-}\right)^2 }
\label{alpha66455}
\end{eqnarray}
and then
\begin{eqnarray}
\beta=\frac{1}{2}\sqrt{-4\alpha^2+f_{\pi}^2}.
\label{be76885}
\end{eqnarray}
For numerical work we use:
\begin{eqnarray}
{m'}_{8^-} &=& 1.30 {\rm GeV},
\nonumber \\
{\rm Tr} \left(Y_8^2\right) &=& \left( 0.980^2 + 1.474^2\right) \,\,\,{\rm GeV^2},
\nonumber \\
\det \left(Y_8^2\right) &=& 0.980^2 \times 1.474^2 \,\,\,{\rm GeV^4}.
\label{res6466377}
\end{eqnarray}

For two choices of $h_0$, the  parameters are given in  Table \ref{T_param_massless}, the predicted masses in Table \ref{T_masses_massless} and the rotation matrices in Table \ref{T_Rotation_matrices_massless}.

\begin{table}[htbp]
	\begin{center}
		\caption{Values of the model parameters in terms of $h_0$ in the massless quarks limit.}
		\begin{tabular}{c|c|c}
			{\rm Parameters} &$h_0=0.80$\,\,${\rm GeV}$&$h_0=1.0$\,\,${\rm GeV}$\\
			\hline
			\hline
			$\alpha$    (${\rm GeV}$)		& 	$4.96 \times 10 ^{-2}$&   $4.96 \times 10 ^{-2}$\\
			$\beta$ (${\rm GeV}$)           &$4.28 \times 10^{-2}$ &  $4.28 \times 10^{-2}$  \\
			$u_1$&$1.93 \times 10 $&$3.01\times 10 $\\
			$u_2$&    $-1.18 \times 10^{3}$&$   -2.99 \times 10^{3}$\\
			$u_3$&$7.57 \times 10^{-1} $&$4.84 \times 10 ^{-1}$\\
			$u_4$&$-3.51$&$-2.81$\\
			$\lambda_2$&$-3.56 \times 10^{-2}$&$-3.65 \times 10^{-2}$\\
			$\lambda_3$&$-2.17\times 10^{-3}$&$-8.87\times 10^{-4}$\\
			$A$ &  0 & 0\\
			\hline
			\hline	
		\end{tabular}
			\label{T_param_massless}
	\end{center}
\end{table}

\begin{table}[htbp]
	\begin{center}
		\caption{Predicted masses in terms of $h_0$ in the massless quarks limit.}
		\begin{tabular}{c|c|c}
			{\rm Masses (GeV)} &$h_0=0.80$\,\,${\rm GeV}$&$h_0=1.0$\,\,${\rm GeV}$\\
			\hline
			\hline
			$m_{0^-}$ 	     & $\approx 0$   &   $\approx 0$ \\
			${m'}_{0^-}$     & 1.60        &  1.60 \\
			$m_{0^+}$        &  $1.81  \times 10^{-4}$ &  $3.10  \times 10^{-4}$\\
			${m'}_{0^+}$     &  1.50              & 1.50 \\
			$m_h$            &  1.60              & 2.00 \\
	     	$m_{8^+}$        &  $9.52  \times 10^{-1}$   &  $9.52  \times 10^{-1}$    \\
	     	${m'}_{8^+}$     &  1.49              &     1.49    \\          	
			\hline
			\hline	
		\end{tabular}
	\label{T_masses_massless}
	\end{center}
\end{table}

\begin{table}[htbp]
	\begin{center}
		\caption{Values of the model parameters in terms of $h_0$ in the massless quarks limit.}
		\begin{tabular}{c|c|c}
			{\rm Masses (GeV)} &$h_0=0.80$\,\,${\rm GeV}$&$h_0=1.0$\,\,${\rm GeV}$\\
			\hline
			\hline
			$\left[{K_{8^-}}\right]^{-1}$     &	
			$\begin{array}{cc}
			0.757   & 0.653 \\
			-0.653  & 0.757
			\end{array}
			$
			&
			$\begin{array}{cc}
			0.757  &   0.653 \\
			-0.653 &   0.757
			\end{array}
			$
			\\
			\hline
			$\left[{{\widehat K}_{0^-}}\right]^{-1}$     &	
			$\begin{array}{cc}
			0.501  &  -0.865 \\
			-0.865 &  -0.501
			\end{array}
			$
			&
			$
			\begin{array}{cc}
			0.501  &  -0.865 \\
			-0.865 &  -0.501
			\end{array}
			$
			\\
			\hline
			$\left[{K_{8^+}}\right]^{-1}$     &	
			$\begin{array}{cc}
			0.216  &  -0.976 \\
			-0.976 &  -0.216
			\end{array}
			$
			&
			$
			\begin{array}{cc}
			0.216   &  -0.976 \\
			-0.976  &  -0.216
			\end{array}
			$
			\\
			\hline
			$\left[{{\widehat K}_{0^+}}\right]^{-1}$     &	
			$\begin{array}{cc}
			0.757  &  0.653 \\
			-0.653 &  0.757
			\end{array}
			$
			&
			$
			\begin{array}{cc}
			0.757  &   0.653 \\
			-0.653 &  0.757
			\end{array}
			$
			\\
			\hline
			\hline	
		\end{tabular}
	\label{T_Rotation_matrices_massless}
	\end{center}
\end{table}

\section*{Acknowledgments}

A.H.F. gratefully acknowledges  the support of College of Arts and Sciences of SUNY Poly in Spring 2018.

\end{document}